\documentclass[amsmath, amssymb,aps, pre,showpacs,
twocolumn
]{revtex4}

\usepackage[utf8]{inputenc}
\usepackage{graphicx}
\usepackage{color}
\usepackage{bigints}
\usepackage{float}
\usepackage{hyperref}

\newcommand{\eps}{\varepsilon}

\DeclareGraphicsExtensions{.pdf}
\graphicspath{{./figs/}}

\begin{document}
\pagestyle{plain}
\title{Bifurcations in models of a society of reasonable contrarians and 
conformists}

\author{Franco Bagnoli}
\email{franco.bagnoli@unifi.it}   
\affiliation{Dipartimento di Fisica e Astronomia,
 Universit\`a di Firenze,\\
 Via G. Sansone 1,  50017 Sesto Fiorentino (FI), Italy; \\also 
 INFN, sez.\ Firenze.}
\author{Ra\'ul Rechtman}
\email{rrs@cie.unam.mx}
\affiliation{Instituto de Energ\'i{}as Renovables, Universidad Nacional 
 Aut\'onoma de M\'exico,\\  Apdo.\ Postal 34, 62580 Temixco Mor., Mexico}

\begin{abstract}
  We study models of a society composed of a mixture of conformist and
  reasonable contrarian agents that at any instant hold one of two
  opinions.  Conformists tend to agree with the average opinion of
  their neighbors and reasonable contrarians to disagree, but revert
 to a conformist behavior in the presence of an overwhelming
  majority, in line with psychological experiments. The model is
  studied in the mean field approximation and on small-world and
  scale-free networks. In the mean field approximation, a large
  fraction of conformists triggers a polarization of the opinions, a
  pitchfork bifurcation, while a majority of reasonable contrarians
  leads to coherent oscillations, with an alternation of
  period-doubling and pitchfork bifurcations up to chaos. Similar
  scenarios are obtained by changing the fraction of long-range
  rewiring and the parameter of scale-free networks related to the
  average connectivity.
\end{abstract}


\maketitle

\section{Introduction}
In a previous work~\cite{bagnoli2013}, we studied the collective
behavior of a society of reasonable contrarian agents. The
rationale was that in some cases, and in particular in the presence of
frustrated situations like in minority games~\cite{minority}, it is
not convenient to always follow the majority, since in this case one
is always on the ``loosing side'' of the market.  This is one of the
main reasons for the emergence of a contrarian attitude. On the other
hand, if all or almost all agents in a market take the same decision,
it is often wise to follow such a trend. We can denote such a
situation with the word ``social norm''. Following an overwhelming
majority is an ecological strategy since it is probable that this
coherent behavior is due to some unknown piece of information, and in
any case the competitive loss is minimal since it equally affects 
the other agents. Indeed, it is well known that in the
presence of an overwhelming majority, individuals tend to align with it,
even if it is in contrast to evidence~\cite{Ash}. 

The agents in our model can hold opinion 0 or opinion
1, representing one of two parties, or against/in favor of an option.
Each agent gathers the average opinion of his
neighbors and changes his opinion according to this average. The time
evolution is synchronous, so the model is essentially a cellular automaton.

Contrarians were introduced in a different socio-physical model by S. Galam~\cite{Galam-Contrarians, Galam-Gemrev}. In this case they had the effect of destroying consensus in a society mainly formed by conformists. In this case, when the fraction of contrarians becomes opinion-dependent,  chaotic dynamics appear~\cite{Galam-chaotic}. 

A society fully composed by reasonable contrarians exhibits interesting
behaviors when changing the topology of the connections. On a
one-dimensional regular lattice, there is no long-range order, the
evolution is disordered and the average opinion is always halfway
between the extreme values 0 and 1. However, adding long-range connections
or rewiring existing ones, we observe the Watts-Strogatz ``small-world'' 
effect, with a transition towards a mean field behavior.  But
since in this case the mean field equation is, for a suitable choice
of parameters, chaotic, we observe the emergence of coherent
oscillations, with a bifurcation cascade eventually leading to a
chaotic-like behavior of the average opinion.  The small-world
transition is essentially a synchronization effect.
Similar effects with a bifurcation diagram resembling that of the logistic map have been observed in a different model of ``adapt if novel - drop if ubiquitous'' behavior, upon changing the connectivity~\cite{Dodds, Harris}.

Since a homogeneous society of unreasonable contrarians is not so reasonable, we study here the problem of collective behaviors in the presence of a mixture of conformists and contrarians. 
To keep things simple, reasonable contrarian and conformist agents
have the same behavior in the presence of an overwhelming majority of
their neighbors.  Conformists become less conformists in the presence
of a large majority. We can call them ``slightly unreasonable''.  The
presence of reasonable contrarians and slightly unreasonable contrarians
avoids  absorbing states which are rather unusual in
real societies. 

In the presence of a strong fraction of conformists, we have the
classical ferromagnetic Ising scenario, with the appearance of a
stationary average opinion different from one half. As this fraction
becomes smaller, this polarized opinion vanishes, as
expected. What is unexpected is  another bifurcation, with 
the appearance of oscillations and chaos as the fraction of conformists
becomes smaller.

The outline of the present paper is as follows: in
Sec.~\ref{sec:model} we present the model in detail. Its mean field
approximation is discussed in Sec.~\ref{sec:meanfield}. Then the model
is studied on small-world networks, Sec.~\ref{sec:smallworld} and on
on scale-free networks, Sec.~\ref{sec:sfn} . We end the presentation
with some conclusions, Sec.~\ref{sec:conclusions}.


\section{The model}
\label{sec:model}  

Our model society is formed by $N$ agents with a fraction $\xi$ of
conformists and a fraction $1-\xi$ of contrarians. Agent $i$,
$i=0,\dots,N-1$ holds an opinion
$s(i,t)\in\{0,1\}$ at time $t$. The opinions of all agents change
synchronously in time. Agent $i$ gathers the average opinion of his
neighbors and changes opinion tending to agree with his neighbors if
he is a conformist, or to disagree if he is a contrarian.

The neighborhoods of all agents are defined
by the adjacency matrix $A$ with components $a_{ij}\,\in \{0,1\}$
in a way that $a_{ij}=1$ if agent $j$ belongs to $i$'s neighborhood and 
$a_{ij}=0$ if he doesn't.  The connectivity $k_i$ of agent $i$ is the number
of agents in $i$'s neighborhood,
\begin{equation}
 k_i=\sum_j a_{ij}
\end{equation}
and the average opinion $h_i$ of his neighbors is  
\begin{equation}
h_i=\dfrac{1}{k_i}\sum_ja_{ij}s(j). 
\end{equation}
The average opinion $c$ of the society is
\begin{equation}
 c=\dfrac{1}{N}\sum_i s_i.
\end{equation} 

\begin{figure}
 \begin{center}
  \includegraphics[width=0.8\columnwidth]{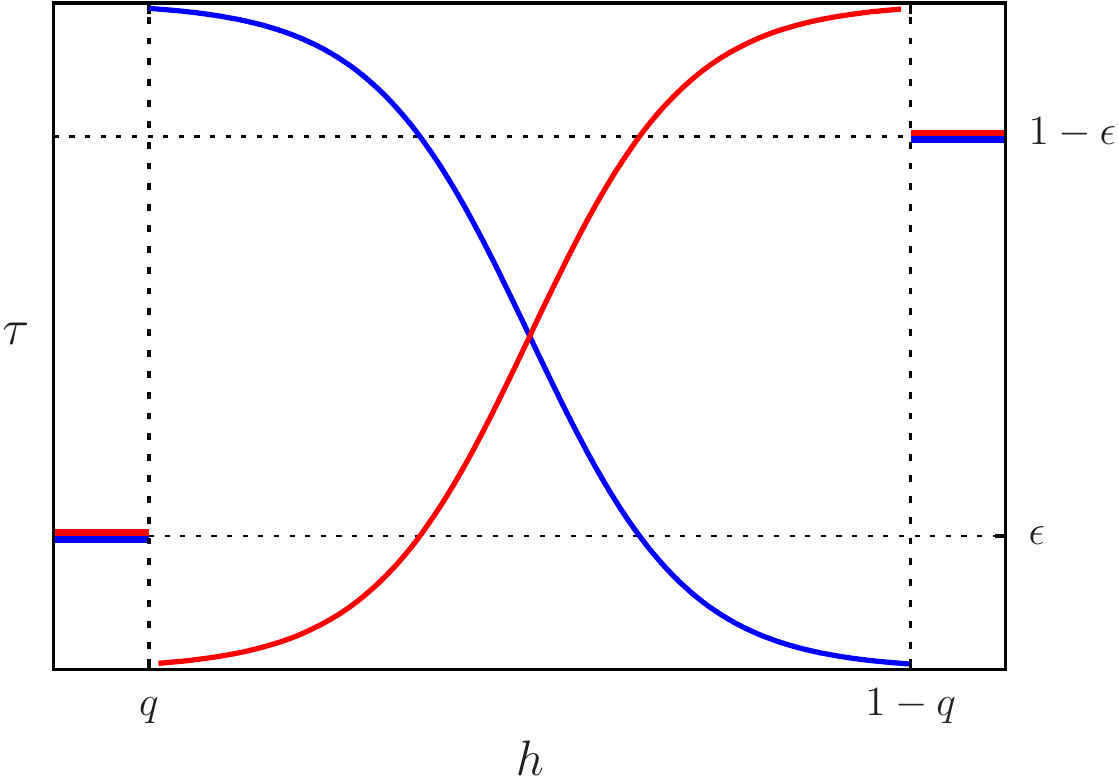}
 \end{center}
 \caption{\label{fig:tau-mf} (Color online) The transition
   probabilities $\tau$ given by Eq.~\eqref{eq:tau} with $|J|=3$. For
   $q<h<1-q$, $\tau$ is an increasing function of $h$ (in red) for
   conformists, $J=3$, and a decreasing one (in blue), $J=-3$, for
   reasonable contrarians. In this and the folowing Figs. $q=0.1$ and
   $\eps=0.2$ unless stated otherwise.}
\end{figure}

Given the average opinion $h$ of the neighbors of agent $i$ at time
$t$, $s(i,t+1)=1$ according to the transition probability $\tau(h;J)$
defined by~\cite{bagnoli2013}
\begin{equation}\label{eq:tau}
 \tau(h;J) =
  \begin{cases}
    \eps & \text{if $h<q$,}\\
    \dfrac{1}{1+\exp(-2J(2h-1))} & \text{if $q\le h\le 1-q$,}\\
    1-\eps & \text{if $h>1-q$.}
  \end{cases}
\end{equation}
In this last expression $J$ represents the agent's conviction,
conformists have $J>0$, reasonable contrarians, $J<0$. The graphs of
$\tau$ are shown in Fig.~\ref{fig:tau-mf}. A reasonable
conformist or a contrarian can assume opinion 1 with probability
$\eps$ ($1-\eps$) when $0\leq h\leq q$ ($1-q\leq h\leq 1$). When
$q<h<1-q$ a reasonable conformist will probably agree with his
neighbors, and a contrarian will probably disagree. Both conformists
and contrarians share the same values of $|J|$, $q$, and $\eps$. A nonzero
value of $\eps$ avoids the presence of absorbing states but causes the
``slightly unreasonable'' behavior of conformists. Unless otherwise
stated, we always use the value $q=0.1$ and $\eps = 0.2$, as in
Ref.~\cite{bagnoli2013}. 

In the mean field approximation $c$ changes deterministically and we
can characterize the properties of its trajectory by means of the
Lyapunov exponent $\lambda$. On the other hand, $c$ changes
probabilistically on networks, so we use Boltzmann's entropy
$\eta$~\cite{Boltzmann,Ott,bagnoli2013}, which is applicable in both
cases. To define $\eta$ we partition the unit interval in $L$ disjoint
equal sized subintervals $I_i$, $i=1,\dots, L$ and find the
probability $q_i$ that $c$ falls in subinterval $I_i$. Then
\begin{equation}
 \eta=\dfrac{-1}{\log L}\sum_{i=1}^L q_i\log q_i.
\end{equation} 
The probabilities $q_i$ are found numerically by finding the frequency
with which an orbit visits each subinterval $I_i$ after a transient.
A fixed point of the trajectory corresponds to $\eta=0$, and when the
orbit visits every subinterval $I_i$ with the same frequency
$\eta=1$. If $L=2^b$ and the orbit is periodic with period $2^a$,
$\eta=a/b$.  In the limit $L\to\infty$, $\eta\to 0$ for periodic
orbits.

For deterministic maps we can compare the behavior of the Lyapunov
exponent $\lambda$ and that of Boltzmann's entropy $\eta$. Values of
$\lambda > 0$ are equivalent to $\eta > 1/2$ when $L$ is sufficiently
large. In other words, deterministic chaos corresponds to $\eta>1/2$
and order to $\eta<1/2$.  By extending this correspondence to the
probabilistic network dynamics, we define disorder whenever $\eta\gtrsim 1/2$
and order when $\eta\lesssim 1/2$.


\section{Mean field approximation}
\label{sec:meanfield}

We start with a model of a society where the neighborhood of each
agent $i$, either conformist or contrarian, is formed by $k$ random
neighbors, i.e., the mean field approximation for a fixed connectivity
$k$. With a fraction $\xi$ of conformists and a fraction $1-\xi$ of
reasonable contrarians, the time evolution of the average opinion $c$
is
\begin{align}
 \label{eq:mf}
  c'=&\sum_{w=0}^k \binom{k}{w} c^w (1-c)^{k-w}\cdot\nonumber\\
     &\qquad  \left[\xi\tau\left(w/k;J\right)+%
       (1-\xi)\tau\left(w/k;-J\right)\right],
\end{align}
with $c$ and $c'$ the average opinions at times $t$ 
and $t+1$ respectively. In the right-hand side
term, the first parenthesis is the $w$ combinations from a set of $k$
elements.

\begin{figure}
 \begin{center}
 \begin{tabular}{cc}
 (a) & (b) \\
 \hskip -2mm
 \includegraphics[width=0.45\columnwidth]{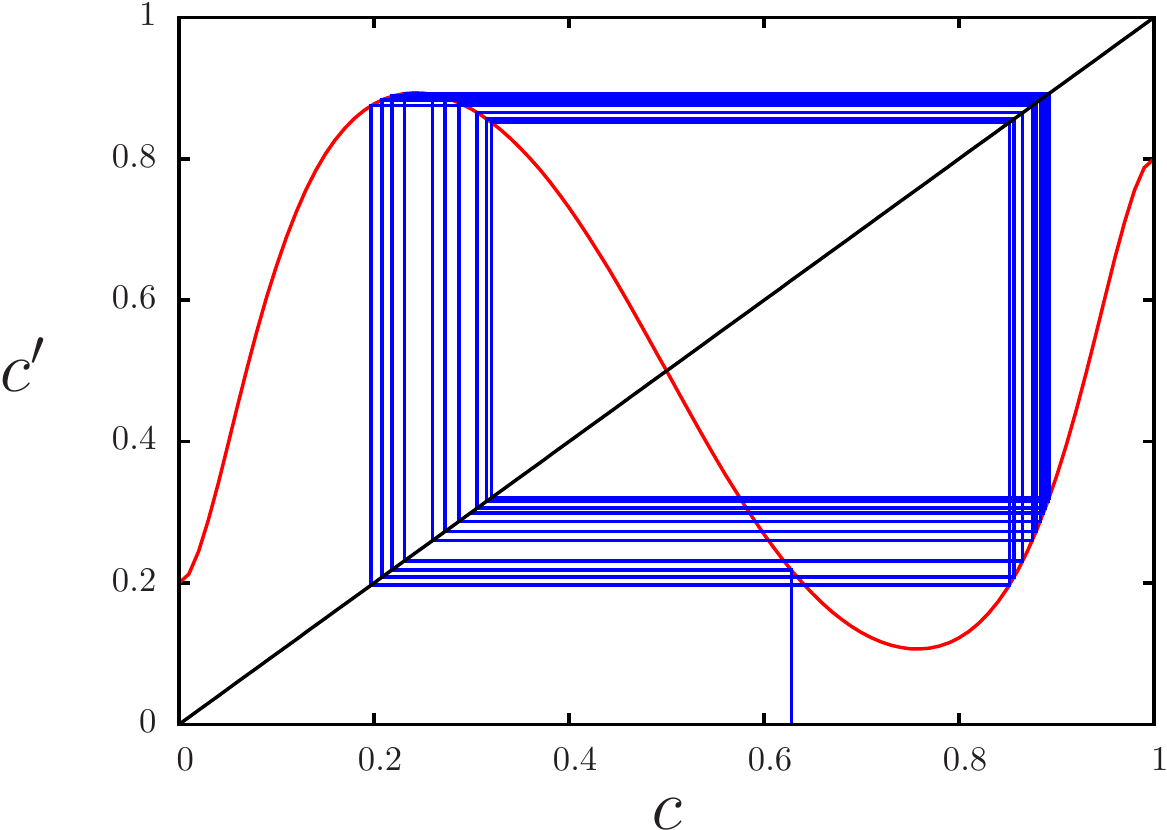} &
 \hskip 2mm
 \includegraphics[width=0.45\columnwidth]{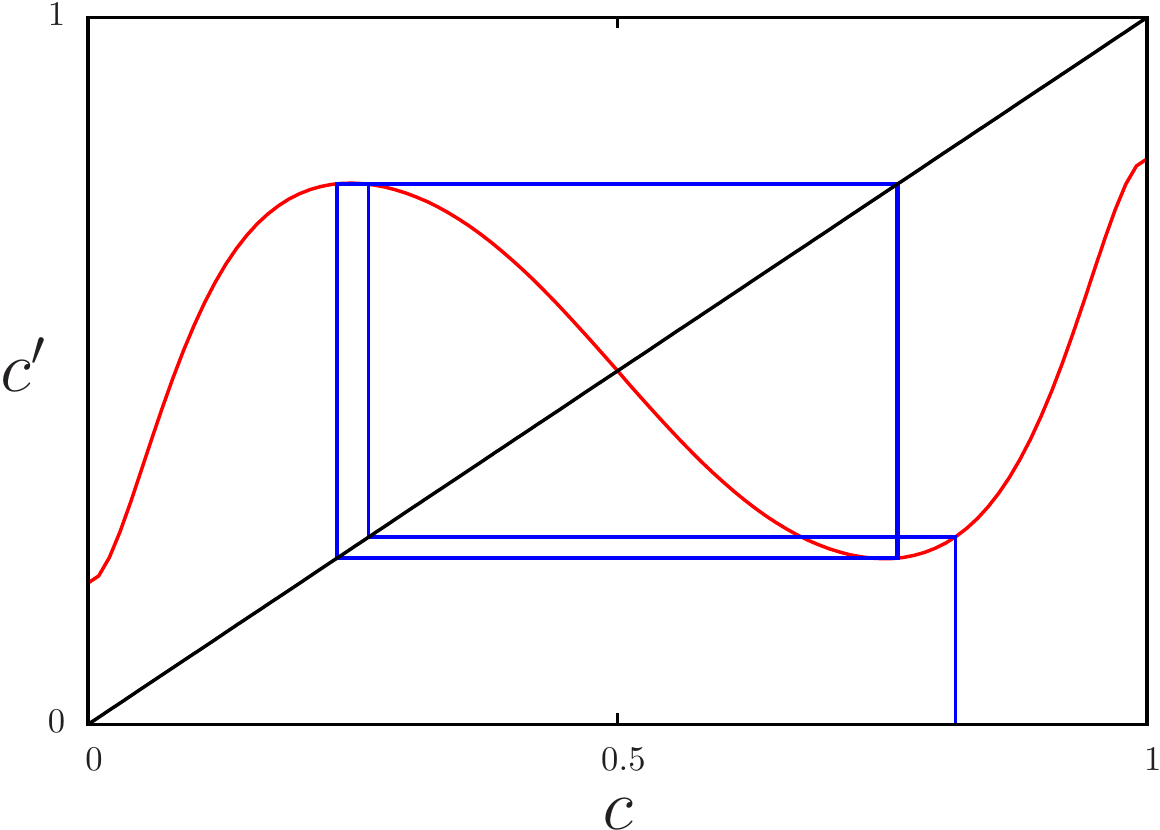} \\
 (c) & (d) \\
 \hskip -2mm
 \includegraphics[width=0.45\columnwidth]{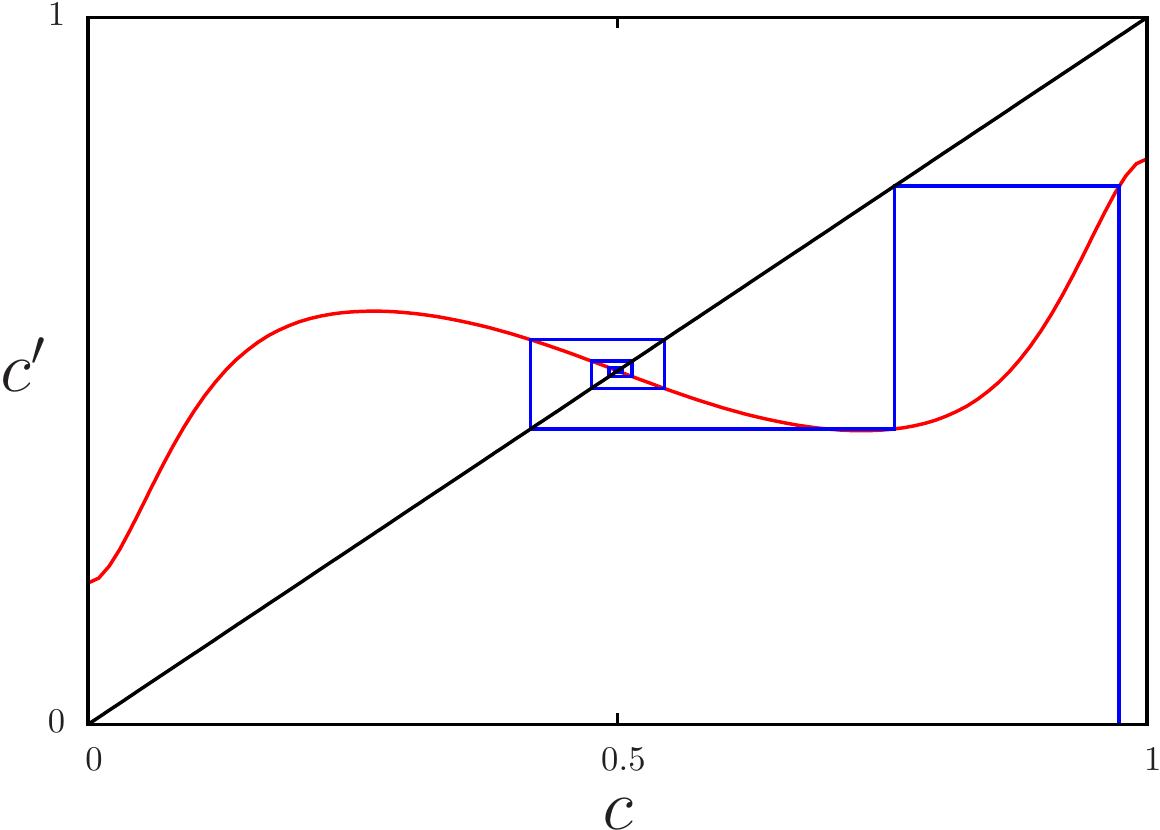} &
 \hskip 2mm
 \includegraphics[width=0.45\columnwidth]{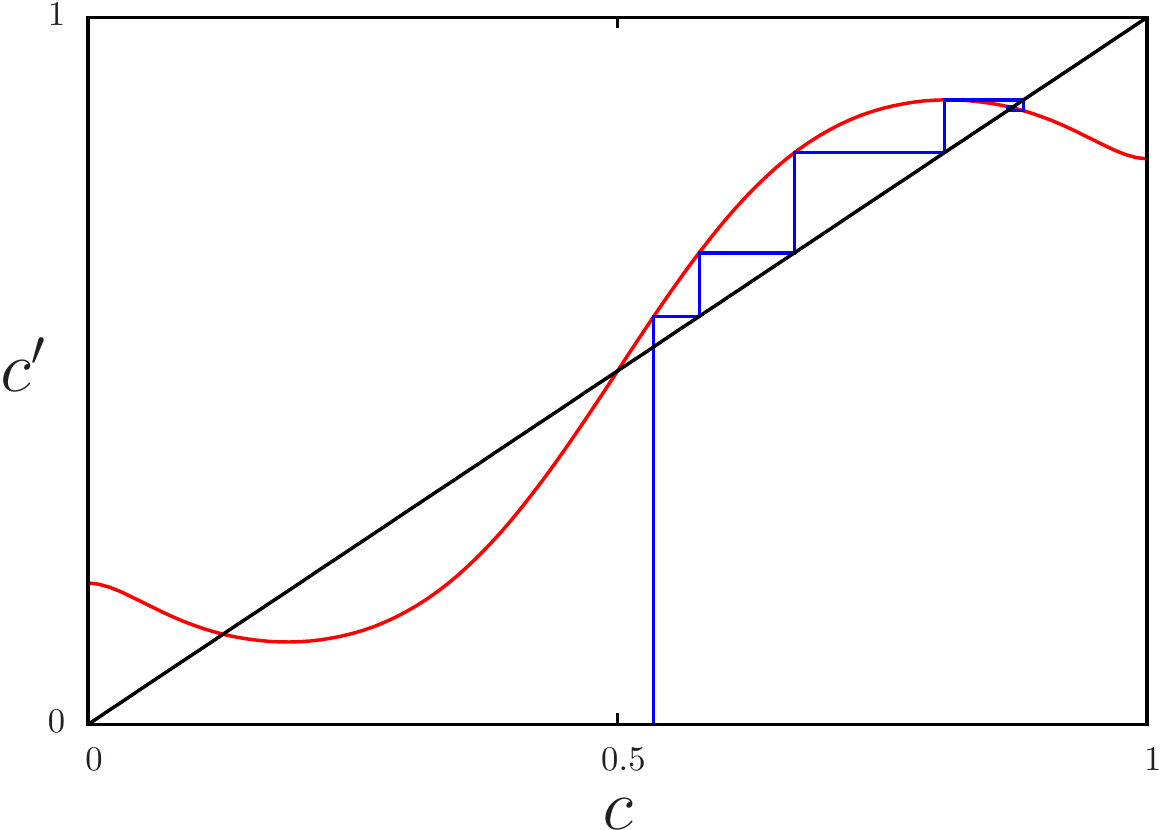} 
 \end{tabular} 
 \end{center}
 \caption{\label{fig:ret-map} (Color online.) Return maps of the mean
   field approximation, Eq.~(\ref{eq:mf}), smooth curves (in red), and
   40 iterates of the map starting from a random initial $c$, stepping
   lines (in blue).  (a) $\xi=0.06$.  (b) $\xi=0.2$. (c)
   $\xi=0.4$. (d) $\xi=0.9$. In this and the following Figs., $|J|=5$,
   $k=20$, $\eps=0.2$, and $q=0.1$ unless stated otherwise.}
\end{figure}

In Fig.~\ref{fig:ret-map} we show return maps of Eq.~(\ref{eq:mf}) for
different values of $\xi$. For small $\xi$, Fig.~\ref{fig:ret-map}
(a), the map is chaotic, and for larger values of $\xi$, we find
periodic orbits or fixed points, Figs.~\ref{fig:ret-map} (b), (c) and
(d).

\begin{figure}
 \begin{center}
 \begin{tabular}{cc}
  (a) & (b) \\
  \hskip -2mm
  \includegraphics[width=0.45\columnwidth]{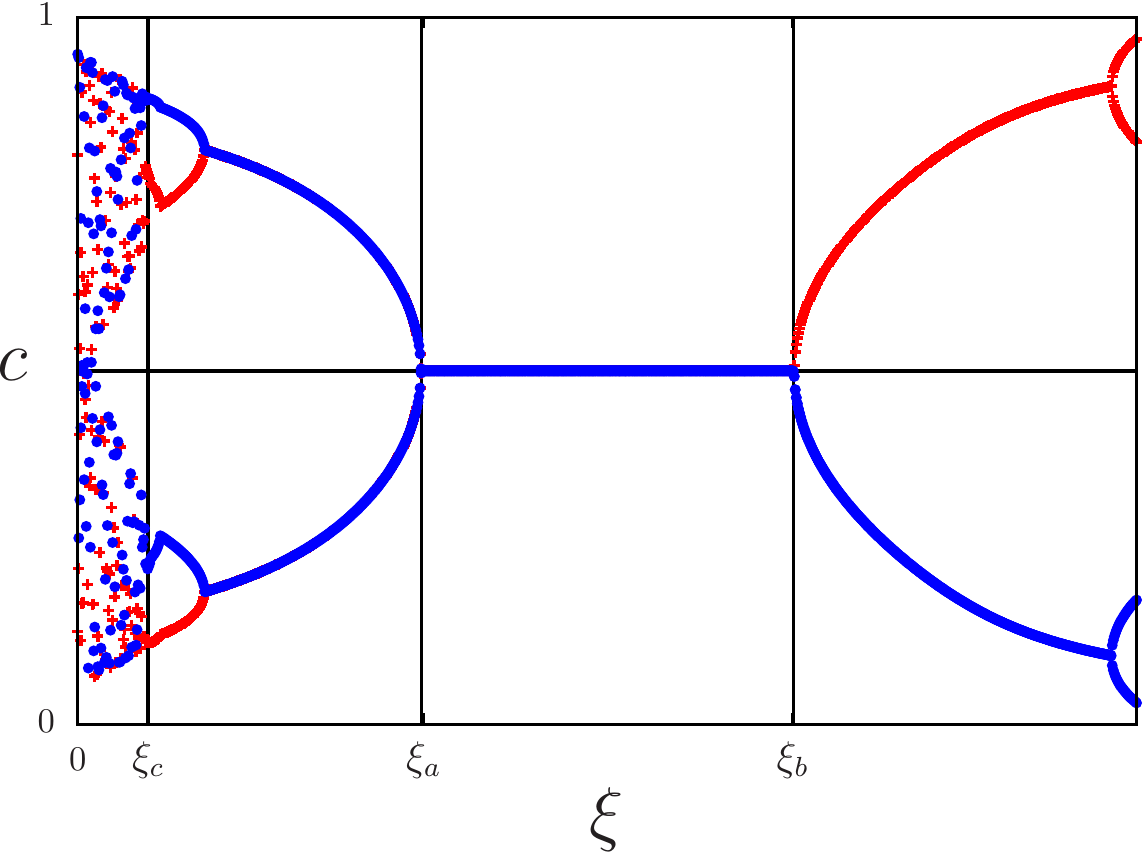} &
  \hskip 2mm\includegraphics[width=0.45\columnwidth]{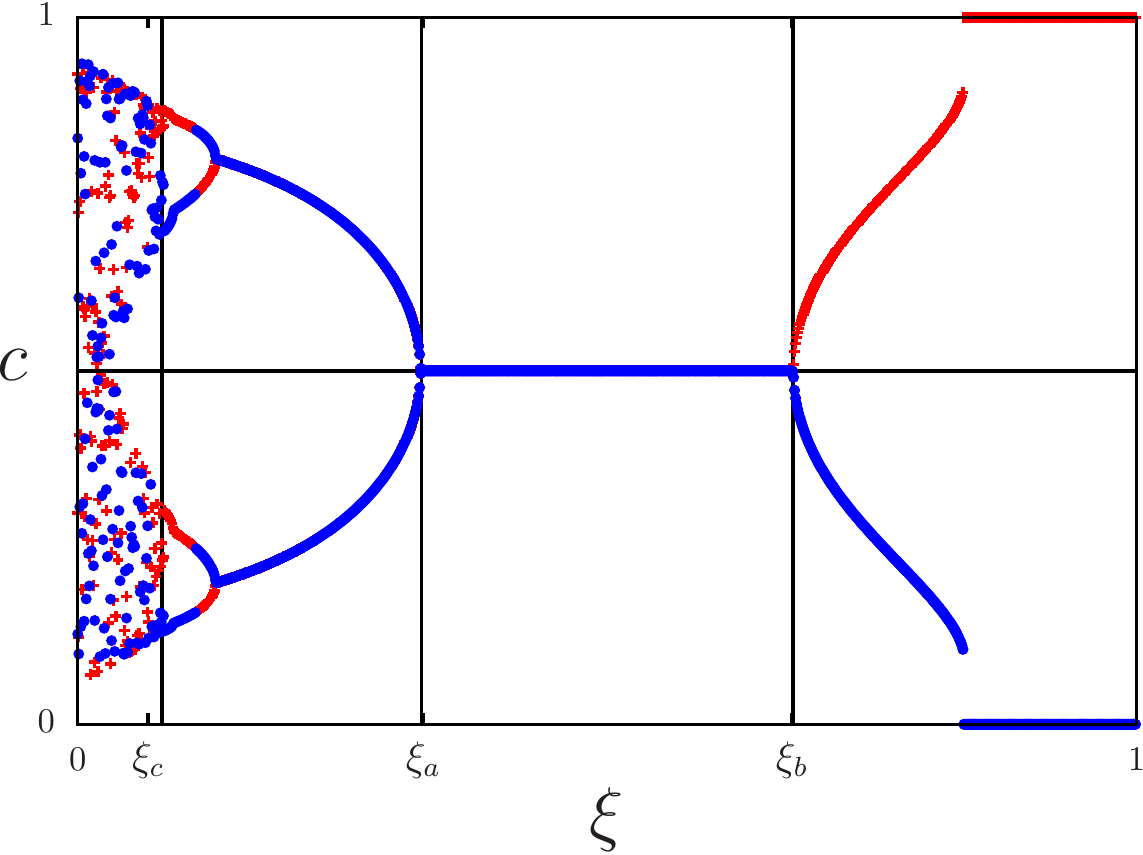} \\
 \end{tabular}
 \caption{\label{fig:c-xi-J5} (Color online.) The mean field average
   opinion $c$ as a function of the fraction of conformists $\xi$ with
   $|J|=5$, $k=20$, and $q=0.1$. (a) $\eps=0.2$, $\xi_c=0.0665$.  (b)
   $\eps=0.0$, $\xi_c=0.0795$.  In (a) and (b), $\xi_a=0.3245$ and
   $\xi_b=0.6755$ so that $1-\xi_b=\xi_a$.  For $\xi$ slightly larger
   than $\xi_c$ the first and third branches (starting from the
   bottom) correspond to $c_0=0.9$ (in red) and the other two to
   $c_0=0.1$ (in blue). For $\xi_b<\xi\leq 1$, the upper branch (in
   red) corresponds to $c_0=0.9$, the lower one (in blue) to
   $c_0=0.1$, For all values of $\xi$, two consecutive iterations are
   plotted after a transient of $1,000$ time steps for both $c_0=0.9$
   (in red) and $c_0=0.1$ (in blue).}
 \end{center}
\end{figure}

\begin{figure}
 \begin{center}
 \begin{tabular}{cc}
  (a) & (b) \\
  \includegraphics[width=0.45\columnwidth]{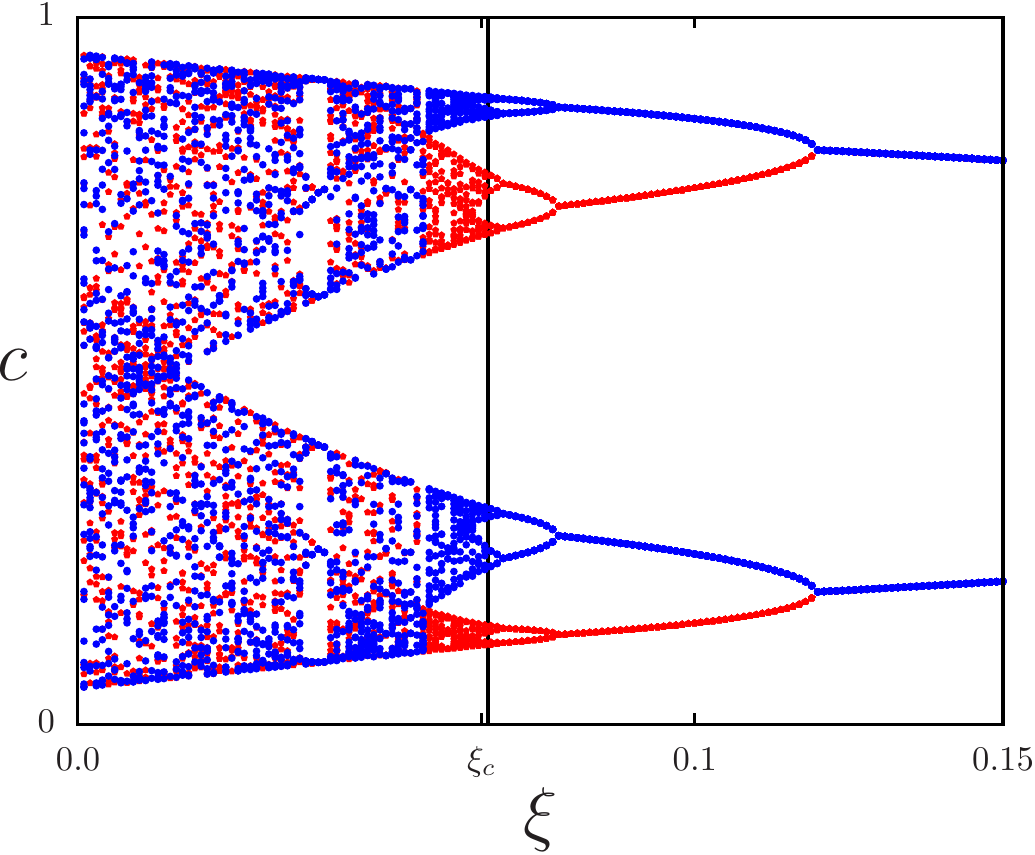} &
  \includegraphics[width=0.5\columnwidth]{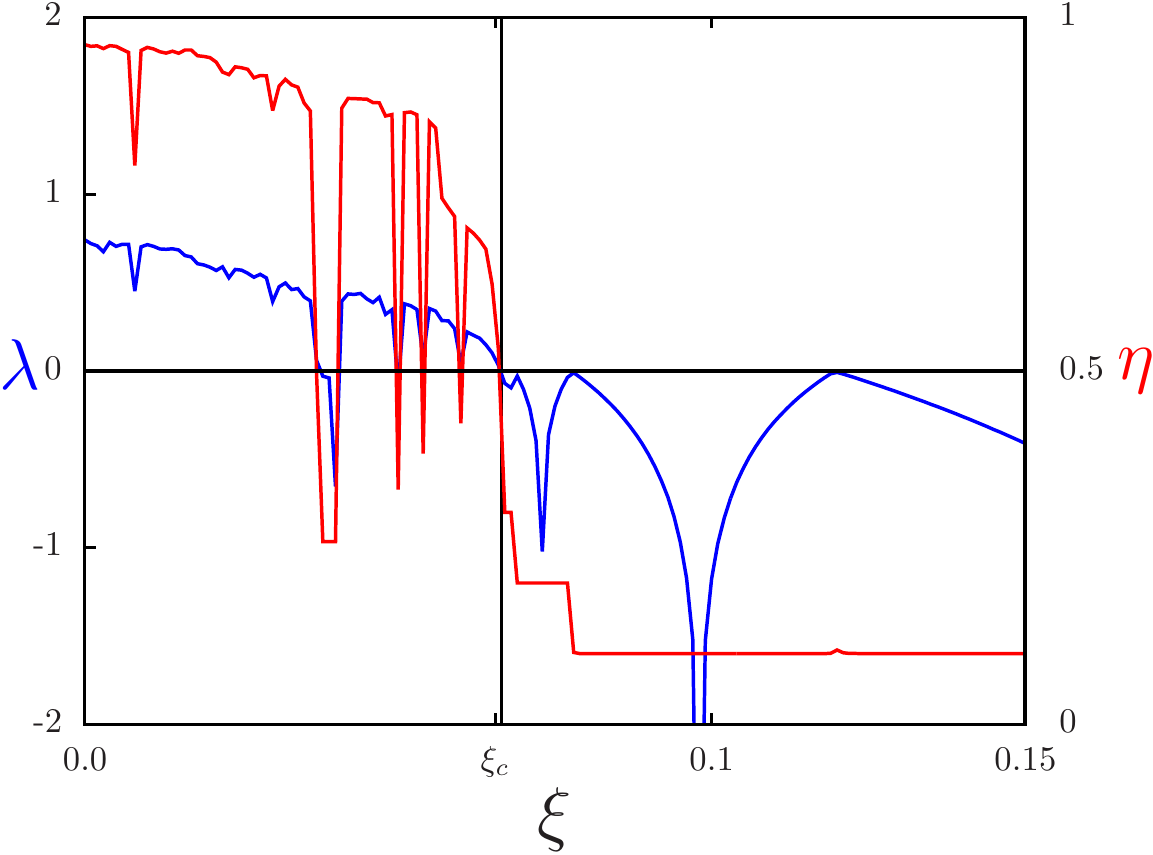}
 \end{tabular}
 \caption{\label{fig:c-lambda-xi-J5} (Color online.) (a) The mean
   field average opinion $c$ as a function of the fraction of conformists $\xi$.
   For
   $\xi$ slightly smaller and also larger than $\xi_c$, we see four
   branches. Starting from below, the first and third branches
   correspond to $c_0=0.9$ (in red), and the second and fourth to
   $c_0=0.1$ (in blue). After a 1,000
   time steps transient, 32 consecutive iterations are plotted for
   each value of $\xi$ starting with $c_0=0.9$ and $c_0=0.1$.
   (b) The Lyapunov exponent $\lambda$, bottom
   curve for small $\xi$ (in blue), and the entropy $\eta$, top
   curve for small $\xi$ (in red), of the mean field average opinion $c$,
   Eq.~(\ref{eq:mf}), as functions of
   $\xi$. For each value of $\xi$,  $\lambda$ is evaluated during 1,000 time steps and 
   for $\eta$, the unit interval is
   divided in $2^{10}=1,024$ equal size subintervals and the frequency
   with which each subinterval is visited is found during $100\times
   2^{10}=102,400$ time steps after a 300 time steps transient.}
 \end{center}
\end{figure}

In Fig.~\ref{fig:c-xi-J5} we show the mean field bifurcation diagrams
of $c$ given by Eq.~(\ref{eq:mf}) as a function of the fraction of
conformists $\xi$ for $\eps=0.2$ and $\eps=0$.  In this second case,
there are absorbing states for large $\xi$ but for smaller values the
diagram hardly depends on the value of $\eps$.  The leftmost vertical
line at $\xi_c$ marks the threshold at which the chaotic region ends.
In Fig.~\ref{fig:c-lambda-xi-J5} (a) we show an amplication of the
bifurcation diagram for small $\xi$ and in
Fig.~\ref{fig:c-lambda-xi-J5} (b) the corresponding Lyapunov exponent
$\lambda$ and Boltzmann's entropy $\eta$. For $\xi>\xi_c$, $\lambda<0$
and $\eta<1/2$.  In Figs.~\ref{fig:c-xi-J5} (a) and (b), the next
vertical line at $\xi_a$ corresponds to the period-doubling
bifurcation from a period two orbit to a fixed point. For
$\xi_c<\xi<\xi_a$ the rather large fraction of contrarians causes
symmetric oscillations of $c$.  The rightmost vertical line at $\xi_b$
corresponds to a pitchfork bifurcation from $c=1/2$ to $c>1/2$ when
$c_0>1/2$ and to $c<1/2$ when $c_0<1/2$ with $c_0$ the average opinion
at $t=0$, (see also Fig.~\ref{fig:ret-map}). 

\begin{figure}
 \includegraphics[width=0.8\columnwidth]{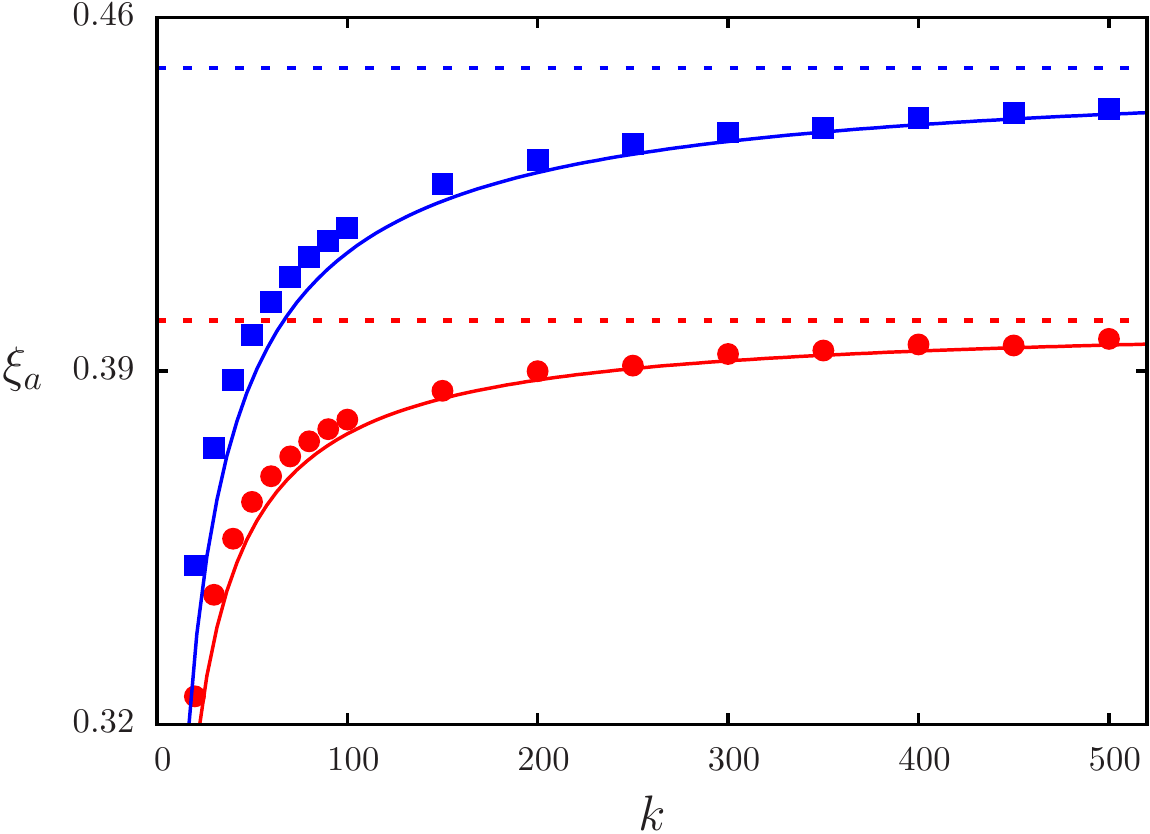}
 \caption{\label{fig:xia} (Color online.) Mean field critical values
   $\xi_a$ as functions of $k$ for $|J|=5$, circles (in red), and
   $|J|=10$ squares (in blue).  The continuous curves are the graphs
   of Eq.~\eqref{xia}, the bottom one (in red) is for $|J|=5$, and the
   top one (in blue) corresponds to $|J|=10$. The horizontal dashed
   lines show the asymptotic values (derived in the Appendix),
   $\xi_a(5, \infty)=0.4$, bottom dashed line (in red), and $\xi_a(10,
   \infty)=0.45$ top dahed line (in blue). Numerical data obtained
   after a transient of 1,000 time steps starting with $c_0=0.9$.}
\end{figure}

We prove in the Appendix that $\xi_a=1-\xi_b$ and find that the approximate
behavior of $\xi_a$ as a function of $J$ and $k$ is given by
\begin{equation}\label{xia}
\xi_a(J, k) = \frac{1}{2}\left(1-\frac{1}{J}\sqrt{1+\frac{2J^2}{k}}\right).
\end{equation}
In Fig.~\ref{fig:xia} we show that Eq.~(\ref{xia}) agrees with the
numerical results for large connectivities $k$.

\begin{figure}
 \begin{center}
 \begin{tabular}{cc}
 (a) & (b) \\
 \hskip -2mm
 \includegraphics[width=0.45\columnwidth]{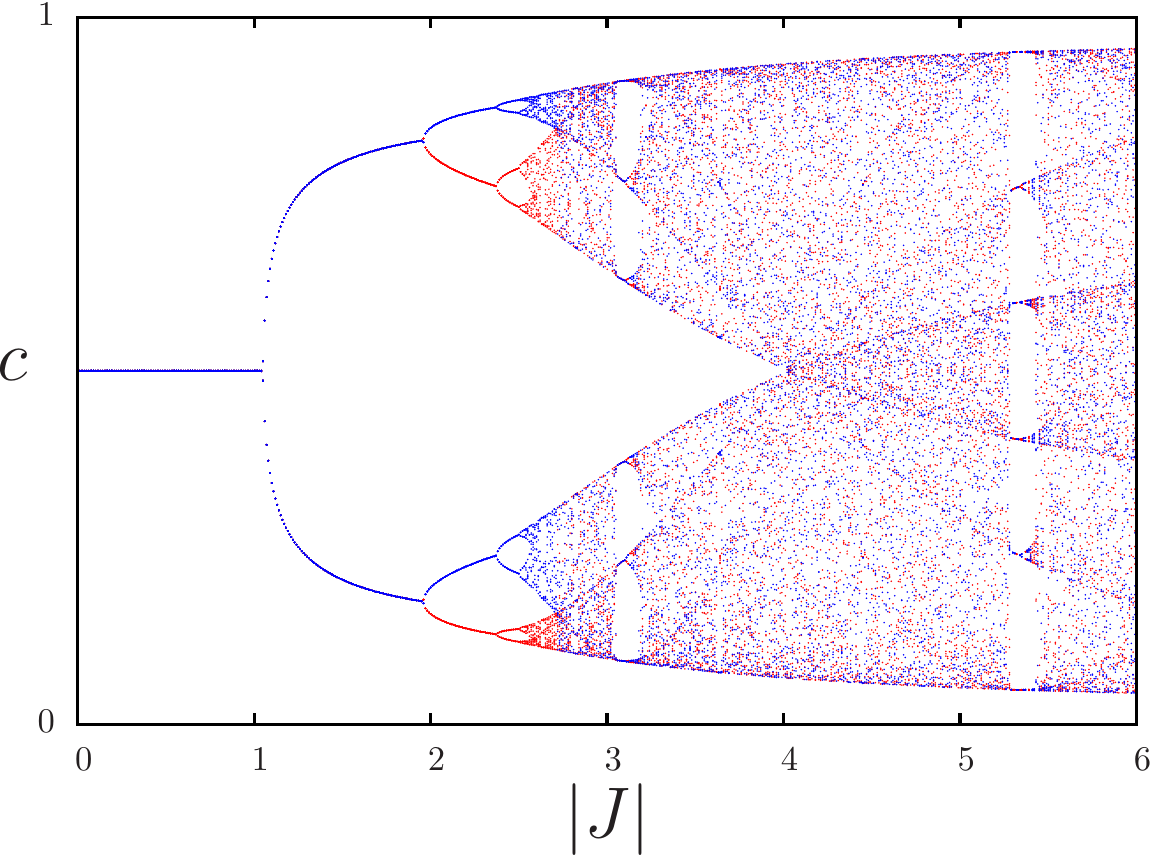} &
 \hskip 2mm
 \includegraphics[width=0.45\columnwidth]{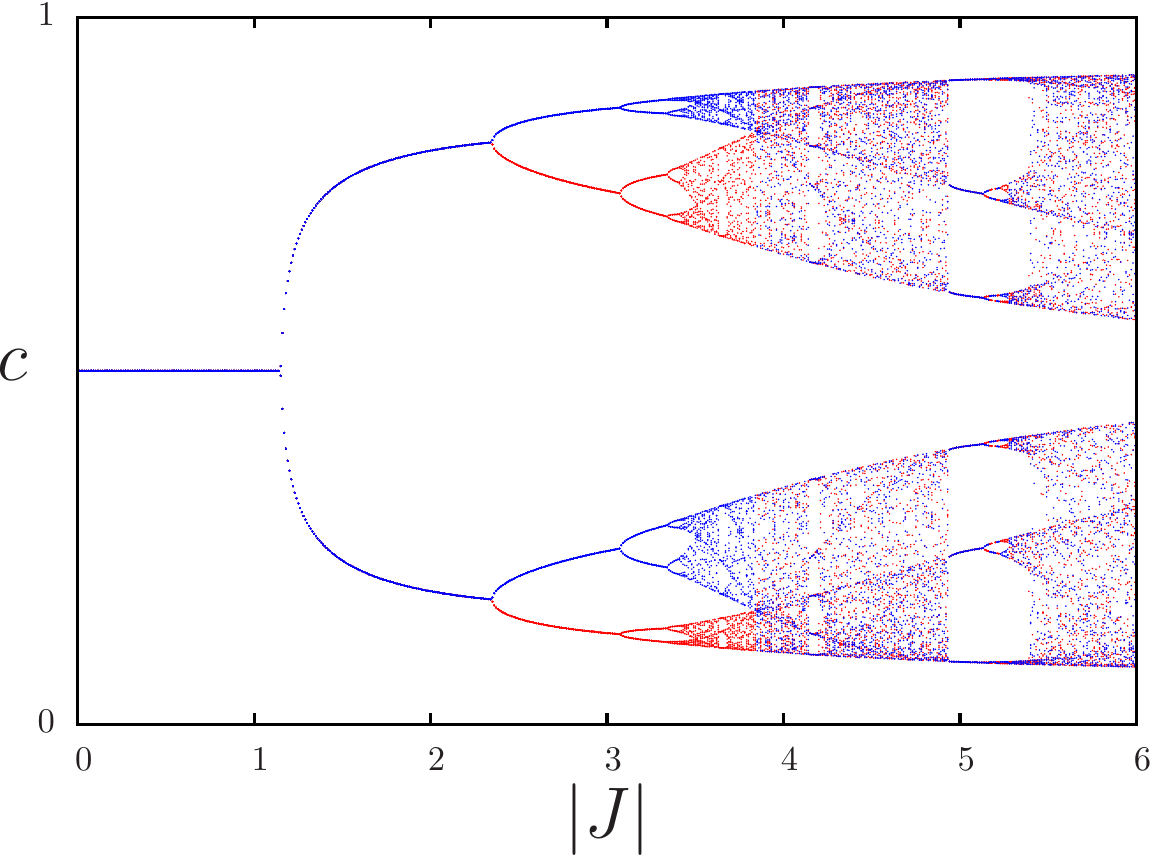} \\
 (c) & (d) \\                          
 \hskip -2mm
 \includegraphics[width=0.45\columnwidth]{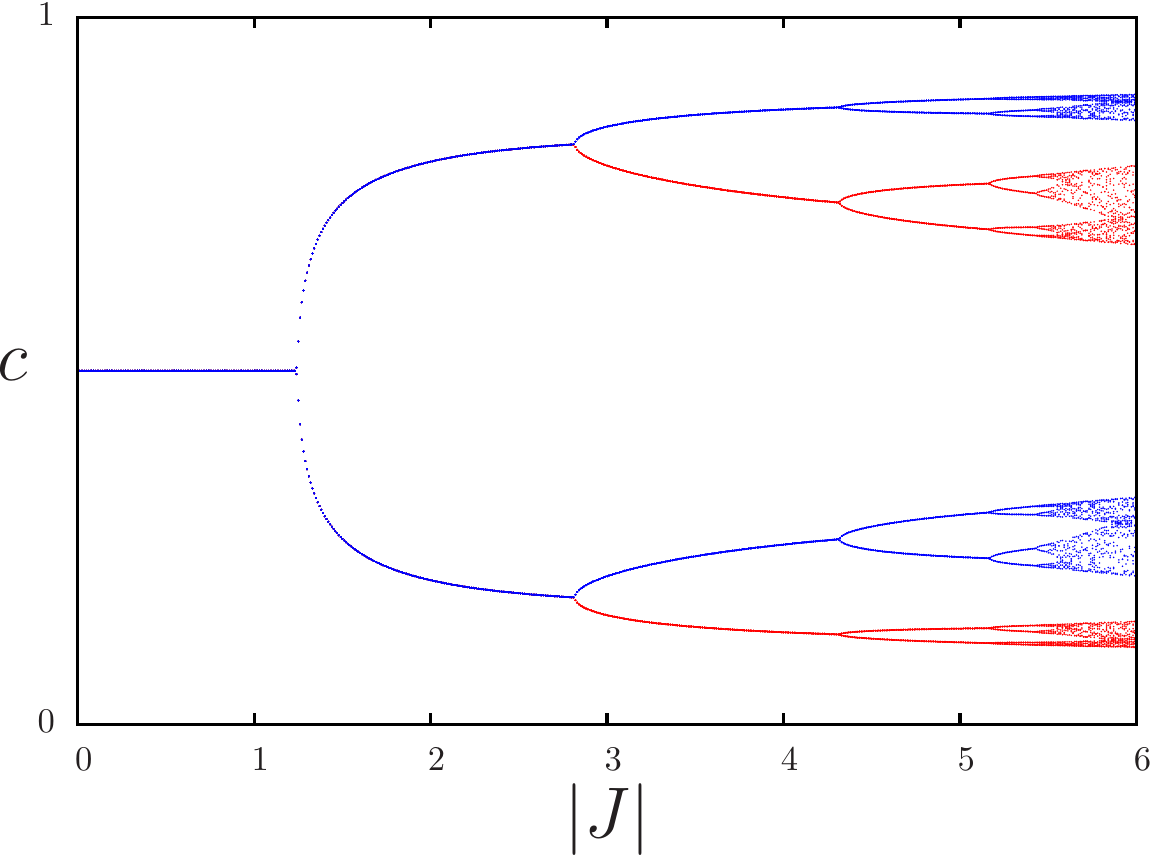} &
 \hskip 2mm
 \includegraphics[width=0.45\columnwidth]{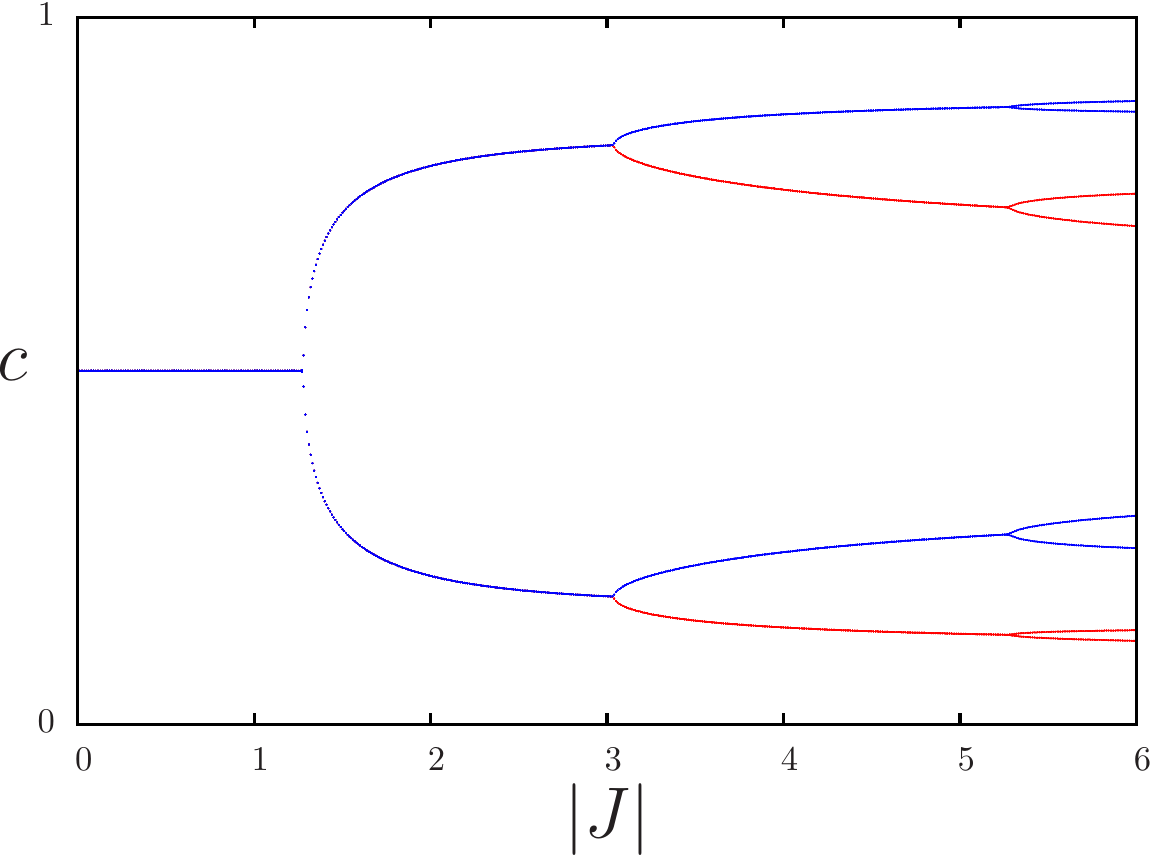} \\
 \end{tabular} 
 \end{center}
 \caption{\label{fig:bif-diag} (Color online.) Mean field bifurcation
   diagrams of the average opinion $c$ as functions of $|J|$ for
   several values of the fraction of conformists $\xi$.  (a)
   $\xi=0.0$. (b) $\xi=0.04$. (c) $\xi=0.07$. (d) $\xi=0.08$.
   Starting from small $|J|$ there is a first bifurcation to a period
   two orbit. For larger values of $|J|$ there are four branches that
   lead. Starting from
   below, the first and third branches are obtained with an initial
   average opinion $c_0=0.9$ (in red), and the second and fourth with
   $c_0=0.1$ (in blue). For each value of $|J|$, after a transient of
   1,000 time steps the next 32 iterations are plotted for two initial
   average opinions $c_0$.}
\end{figure}

In Fig.~\ref{fig:bif-diag}, we show bifurcation diagrams of the
average opinion $c$ as $|J|$ changes for small values of $\xi$. For
$\xi=0$, Fig.~\ref{fig:bif-diag} (a), there are no conformists and we have the
case studied in Ref.~\cite{bagnoli2013}. The bifurcation diagram 
is stretched horizontally as $\xi$ grows. For small $|J|$, $c=1/2$, and as this
parameter grows, there is a first bifurcation to a period 2
orbit. What seems like a second bifurcation to a period-four orbit,
corresponds to two period-two bifurcations that depend on the initial
opinion $c_0$. Starting from below, the first and third branches (in
red) correspond to $c_0=0.9$, and the other two (in blue) to
$c_0=0.1$.  


\section{Small-world networks}
\label{sec:smallworld}

\begin{figure}
 \begin{center} 
 \begin{tabular}{cc}
 (a) & (b) \\
 \hskip -2mm
 \includegraphics[width=0.45\columnwidth]%
             {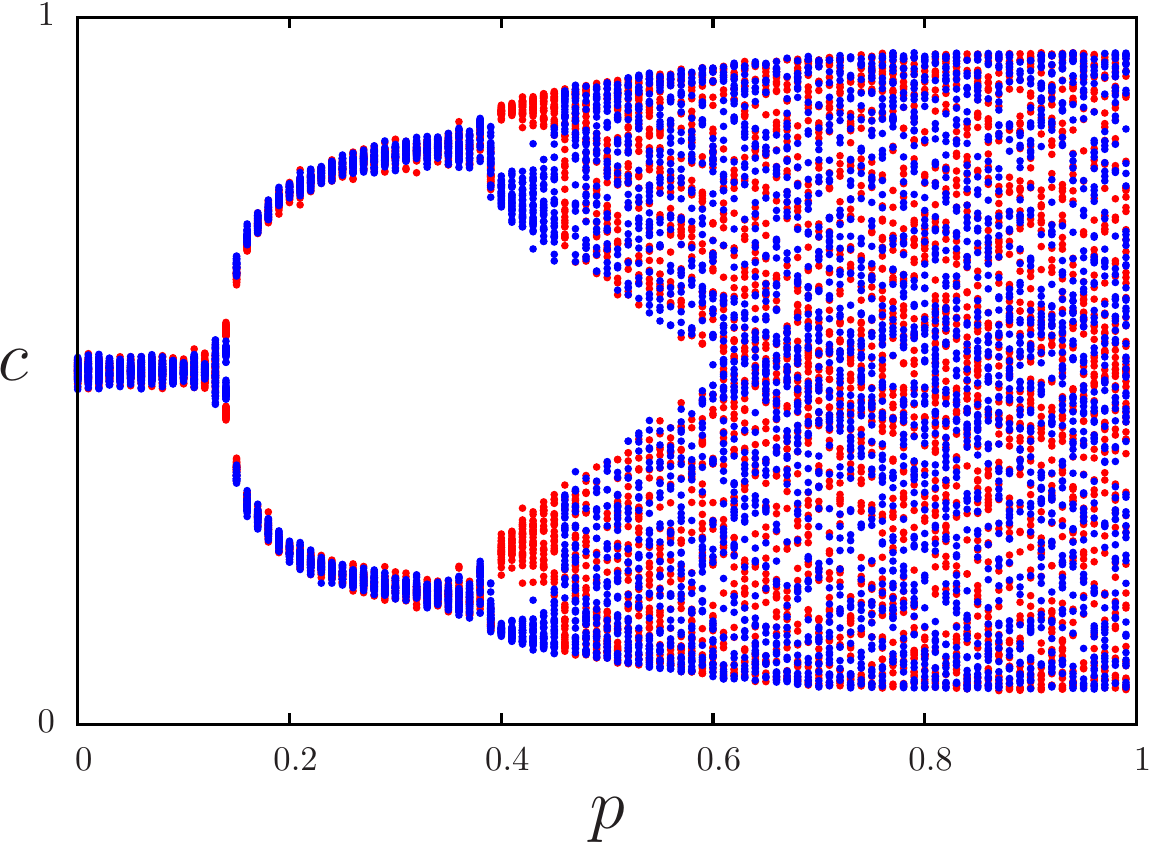} &
 \hskip 2mm
 \includegraphics[width=0.45\columnwidth]%
             {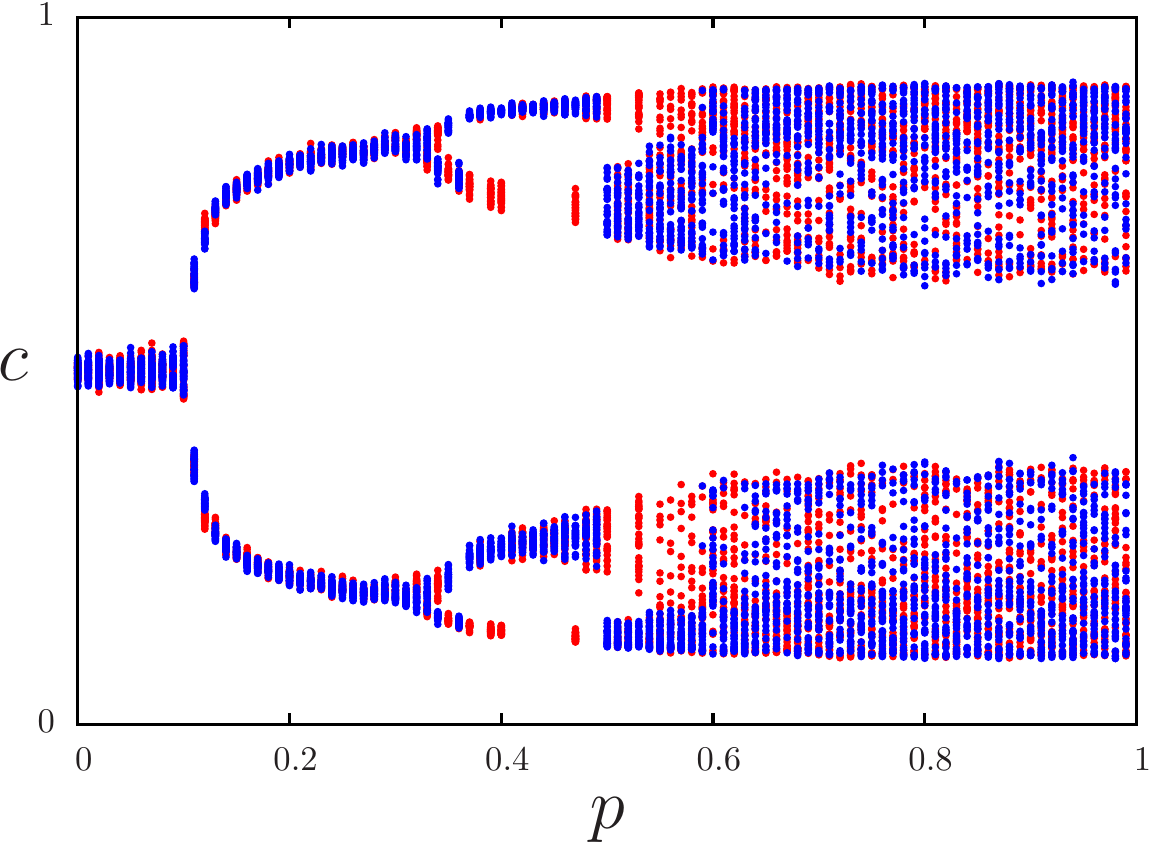} \\
 (c) & (d) \\
 \hskip -2mm
 \includegraphics[width=0.45\columnwidth]%
             {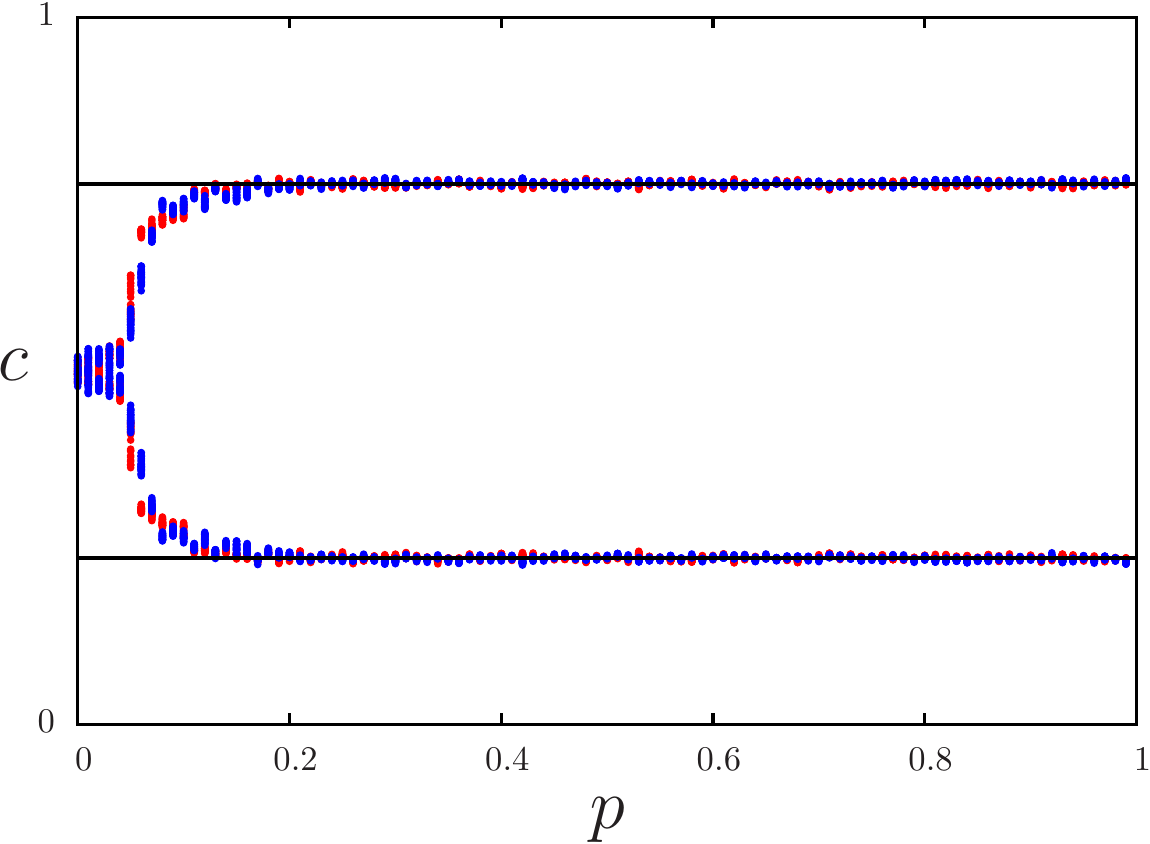} &
 \hskip 2mm
 \includegraphics[width=0.45\columnwidth]%
             {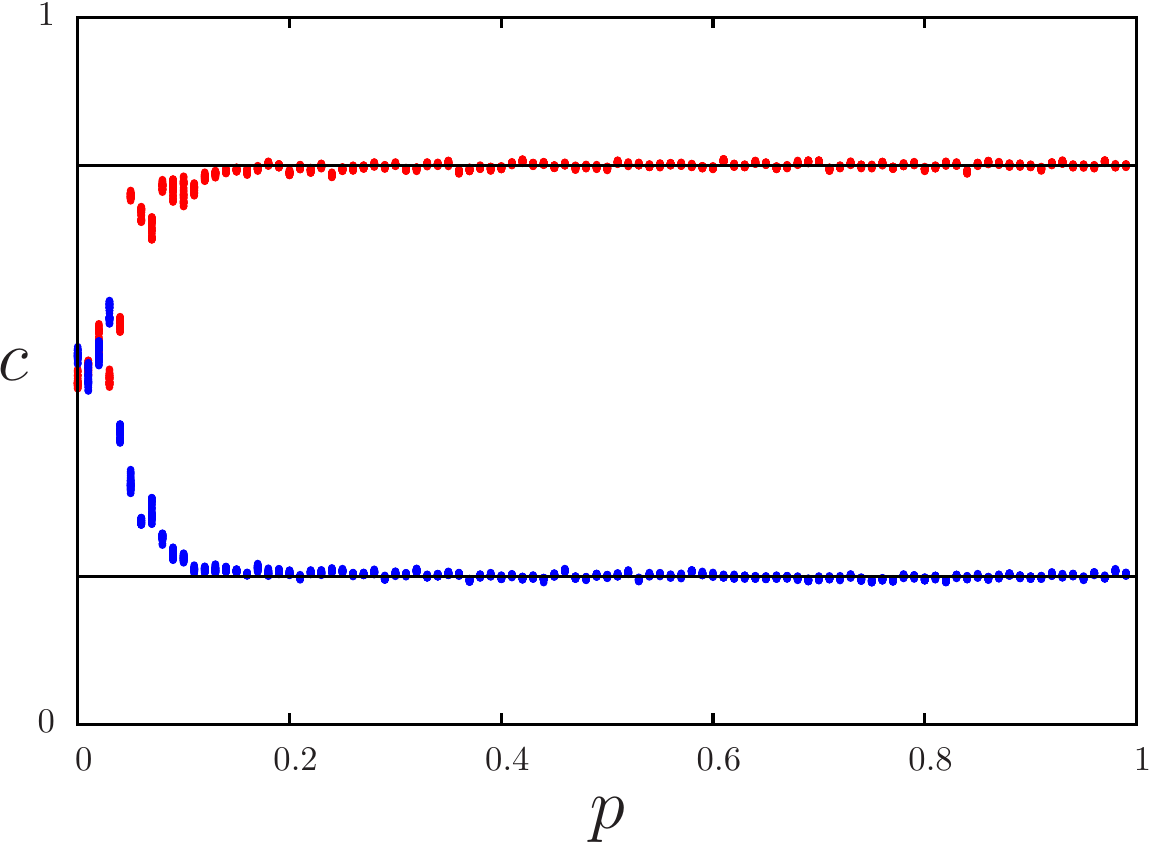} 
 \end{tabular} 
 \end{center}
 \caption{\label{fig:sw-bif-diag} (Color online.) Bifurcation diagrams
   of the average opinion $c$ on a small-world network as a function
   of the long range connection probability $p$ for different values
   of the fraction of conformists $\xi$ with $N=20,000$, $|J|=5$,
   $k=20$. The initial opinion of each agent is chosen at random
   between 0 and 1 with $c_0=0.1$ (blue points) and $c_0=0.9$ (red
   points). (a) $\xi=0.0$.  (b) $\xi=0.05$. (c) $\xi=0.20$.  (d)
   $\xi=0.8$.  In (c) and (d) the horizontal lines are the values of
   $c$ of the period two orbit of Eq.~(\ref{eq:mf}). For each value of
   $p$, 32 points are plotted after a 1,000 time step transient for each
   $c_0$.}
\end{figure}

Real societies are not random, nor regular. It is interesting to study
what happens when the topology changes, due for instance to advances
in the transportation system, or to politics favoring mixing, etc. We
studied the effect of rewiring a fraction $p$ of links in a regular,
one-dimensional society with connectivity $k$. This leads to the
small-world networks first discussed by Watts and
Strogatz~\cite{WattsStrogatz}.  As $p$, the long range connection
probability, grows, small-world networks approach a mean field
behavior. As we show in Fig.~\ref{fig:sw-bif-diag}, the bifurcation
diagrams of $c$ as functions of $p$ are similar to those obtained by
varying $J$ in the mean field approximation.
 
In Fig.~\ref{fig:sw-bif-diag} (a) we study a society of reasonable
contrarians. For small values of $p$, $c$ fluctuates around
$c=0.5$. For slightly larger values, there are noisy oscillations
around two symmetric values, in a way reminiscent of the
period-doubling bifurcations of deterministic systems. For even larger
values, $p\simeq 0.4$, we observe the appearence of two
different noisy oscillating states of period 2. For $c_0=0.1$ the
orbit oscillates between the first and third branches, starting from
the bottom (blue points), and for $c_0=0.9$ the orbit oscillates
between the second and fourth branches (red points).  This roughly
corresponds to what is shown in Fig.~\ref{fig:bif-diag} (a), although
in that case the diagram is drawn as a function of $|J|$ and here as a
function of $p$. For $\xi=0.05$, Fig.~\ref{fig:sw-bif-diag} (b) we
have a similar scenario and for larger values of $\xi$,
Figs.~\ref{fig:sw-bif-diag} (c) and (d), we find a great similarity
with the mean field behavior. 

For $\xi=0.2$, Fig.~\ref{fig:sw-bif-diag} (c), $c$ fluctuates around
$c=0.5$ for small values of $p$ and fluctuates around two branches for
larger values. These branches agree with the period 2 orbit of the
mean field approximation, Eq.~(\ref{eq:mf}) shown in
Fig.~\ref{fig:c-xi-J5}.  For $\xi=0.8$, Fig.~\ref{fig:sw-bif-diag}
(d), and $p\gtrsim 0.2$, $c$ fluctuates around one of two values,
depending on $c_0$, the average opinion at $t=0$. These values also
agree with the mean field approximation shown in
Fig.~\ref{fig:c-xi-J5}.

It is possible to roughly understand these
results assuming that the main contributions to the mean field
character of the collective behavior come from the fraction of links
that are rewired (long-range connections)  that depend on $p$.
The actual value of the field $(2h-1)$ in Eq.~\eqref{eq:tau} is
multiplied by a factor $p$, so that changing $p$ is roughly equivalent
to changing $J$.

\begin{figure}[h]
 \begin{center} 
 \begin{tabular}{cc}
 (a) & (b) \\
 \hskip -2mm
 \includegraphics[width=0.45\columnwidth]%
             {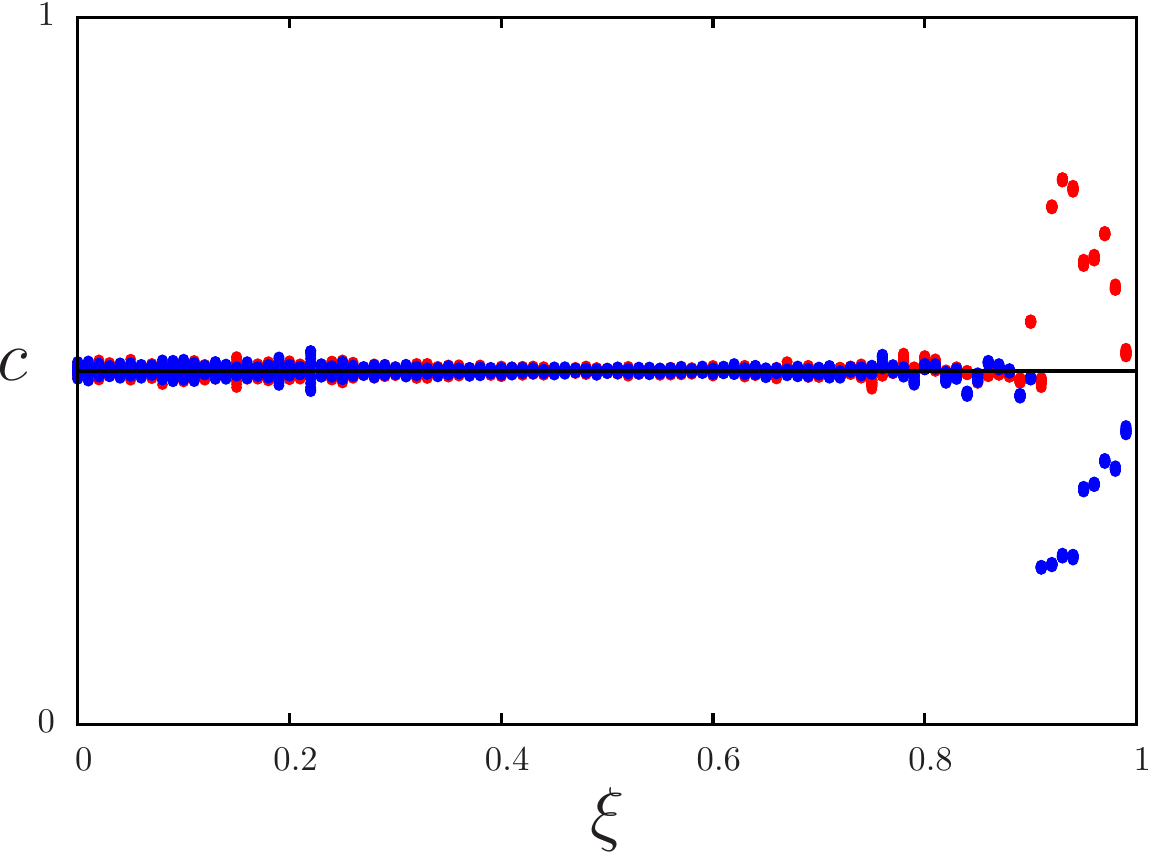} \hskip 5mm&
 \hskip 2mm
 \includegraphics[width=0.45\columnwidth]%
             {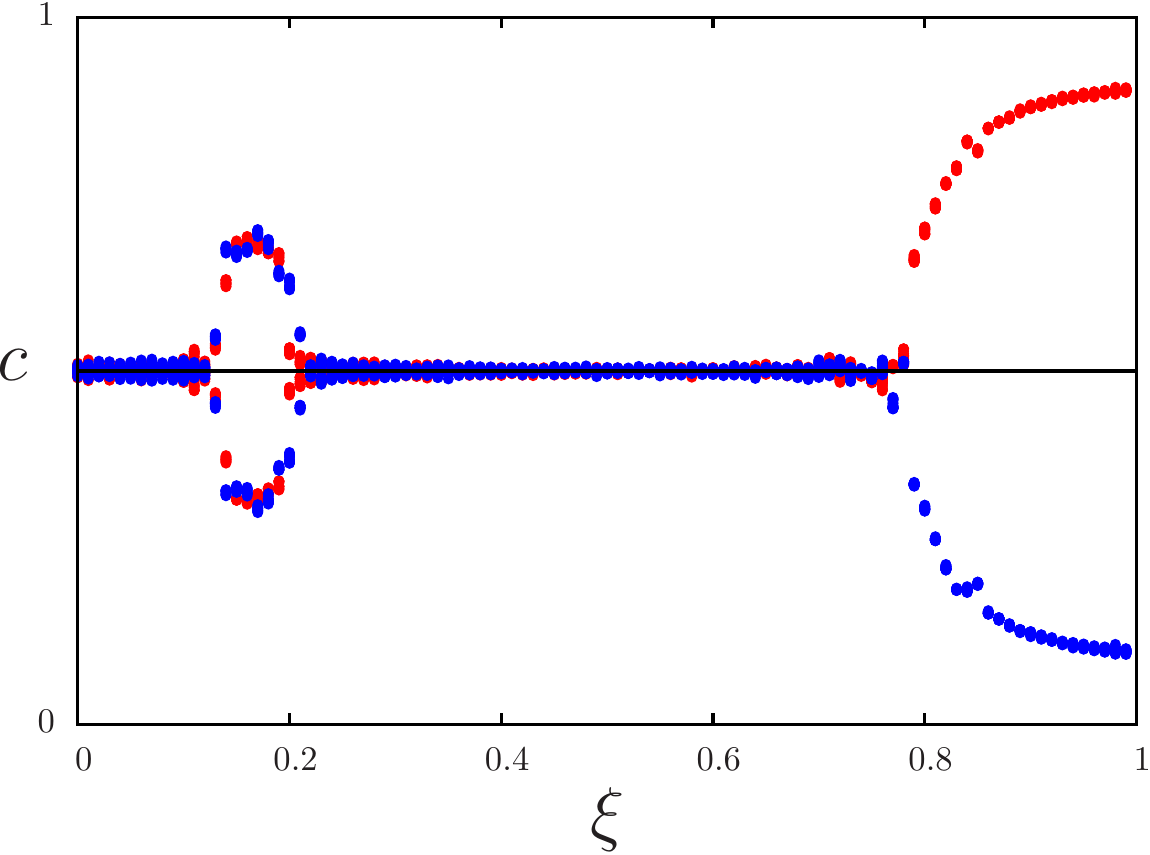} \\
 (c) & (d) \\
 \hskip -2mm
 \includegraphics[width=0.45\columnwidth]%
             {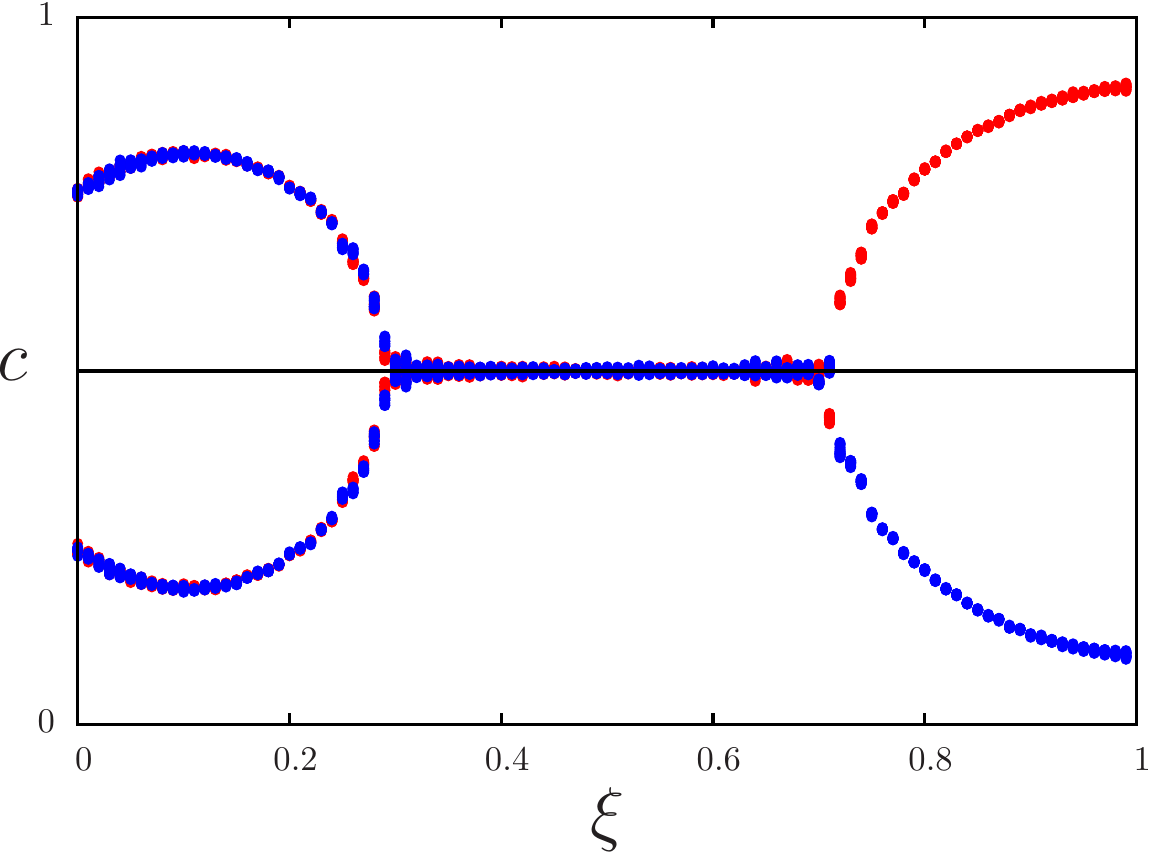} \hskip 5mm&
 \hskip 2mm
 \includegraphics[width=0.45\columnwidth]%
             {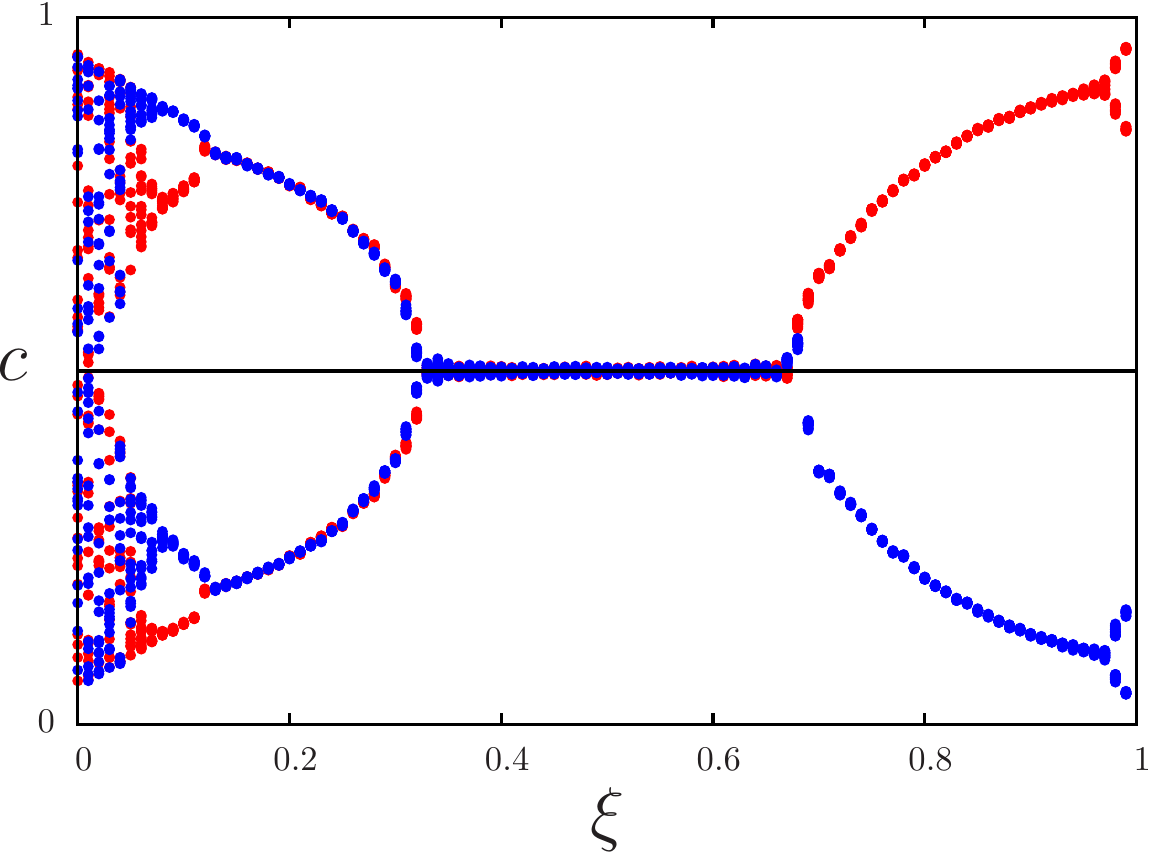} 
 \end{tabular} 
 \end{center}
 \caption{\label{fig:sw-c-xi} (Color online.) Bifurcation diagrams of
   the average opinion $c$ on a
   small-world network as functions of the fraction of conformists
   $\xi$ for different values of the long range connection probability
   $p$ with $N=50,000$, $|J|=5$, $k=20$.
   (a) $p=0.0$. (b) $p=0.05$. (c) $p=0.20$. (d) $p=0.8$. For each
   value of $p$, at $t=0$ the opinion of each agent is chosen at
   random in such a way that the average opinion is $c_0=0.1$ or
   $c_0=0.9$. After a 50,0000 time step transient, 32 points are
   plotted. For $0.6\lesssim\xi\lesssim 0.8$, the orbits that started
   with $c_0=0.99$ have $c>1/2$, (points in red), and those that started
   with $c_0=0.1$ have $c<1/2$, (points in blue).}
\end{figure}

In Fig.~\ref{fig:sw-c-xi} we show bifurcation diagrams of $c$ on
small-world networks as a function of the fraction of conformists
$\xi$ for some values of the long range connection probability
$p$. This sequence of plots illustrate an unexpected behavior.  For
$p=0$, Fig.~\ref{fig:sw-c-xi} (a), there is a pitchfork bifurcation at
$\xi\sim 0.9$.  For larger values of $p$, the pitchfork bifurcation
occurs at smaller values of $\xi$ and an oscillating ``bubble'' is
formed, Fig.~\ref{fig:sw-c-xi} (b) for $0.12\lesssim\xi\lesssim
0.22$. This oscillating region grows with $c$ and for $p=0.8$,
Fig.~\ref{fig:sw-c-xi} (d), the bifurcation diagram is similar to the
mean field one, Fig.~\ref{fig:c-xi-J5} (a).

We can explain this behavior assuming that the conformist behavior promotes synchronization, as does $p$. So the system progressively synchronizes starting from high values of $\xi$, but this synchronization is not visible if the dynamics leads to fixed points. When the synchronization reaches the oscillating phases, it becomes manifest by means of the coherent oscillation of the population. So, this coherent dynamical behavior appears to start first in the vicinity of the bifurcation for $\xi\simeq 0.2$ and then, by increasing $p$, it propagates to lower values of $\xi$. 

\section{Scale-free networks}
\label{sec:sfn}

In this section we present results of the model on uncorrelated
scale-free networks~\cite{Barabasi}. Starting from a fully connected
group of $m$ agents, other $N-m$ agents join sequentially, each one
choosing $m$ neighbors among those already in the group. The choice is
preferential, the probability that a new member chooses agent $i$ is
proportional to its connectivity $k_i$, the number of neighboring
agents that agent $i$ already has. Another way of building the network
is choosing a random edge of a random node and connecting to the other
end of the edge, since such an edge arrives to a vertex with
probability proportional to $kp(k$) with $p(k)$ the probability that a
randomly selected node has connectivity
$k$~\cite{newman-PRE-64-026118-2001}.

\begin{figure}[h]
 \begin{center} 
 \begin{tabular}{cc} 
  (a) & (b) \\
 \hskip 0mm\includegraphics[width=0.45\columnwidth]{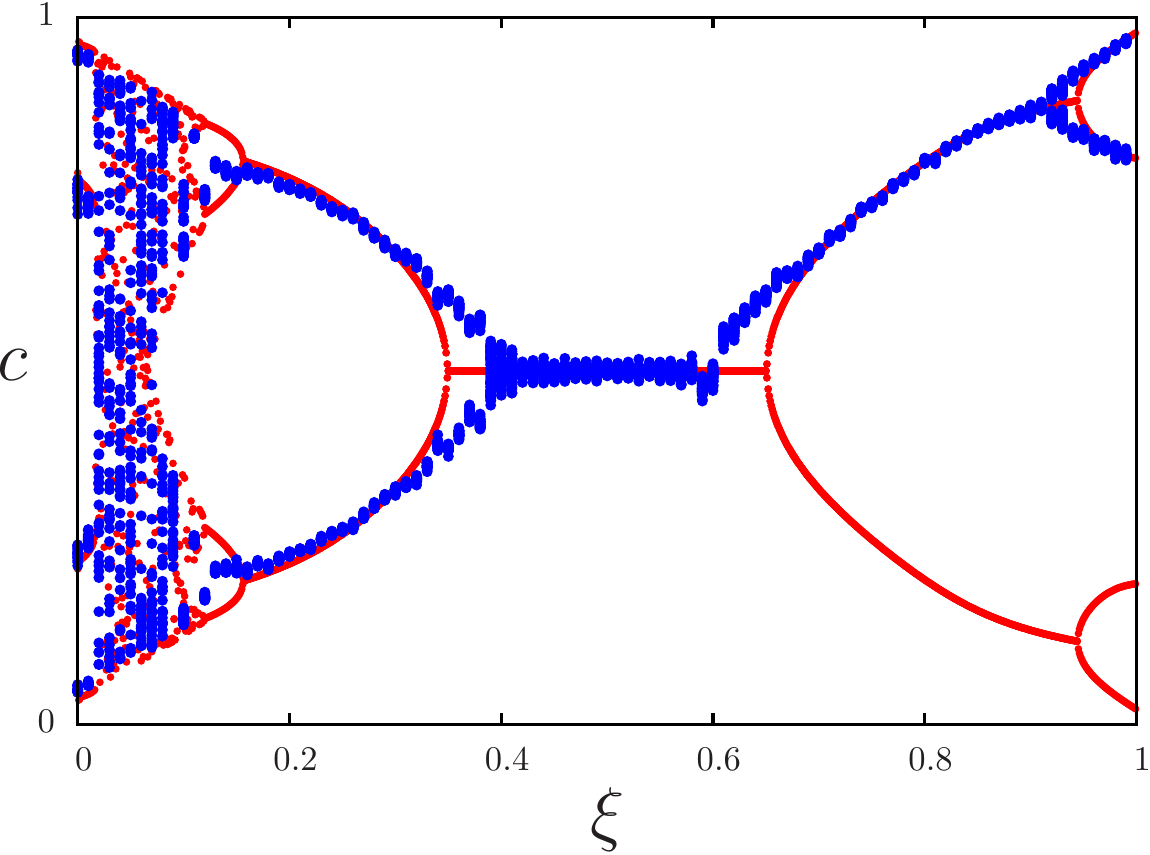} &
 \hskip 2mm\includegraphics[width=0.45\columnwidth]{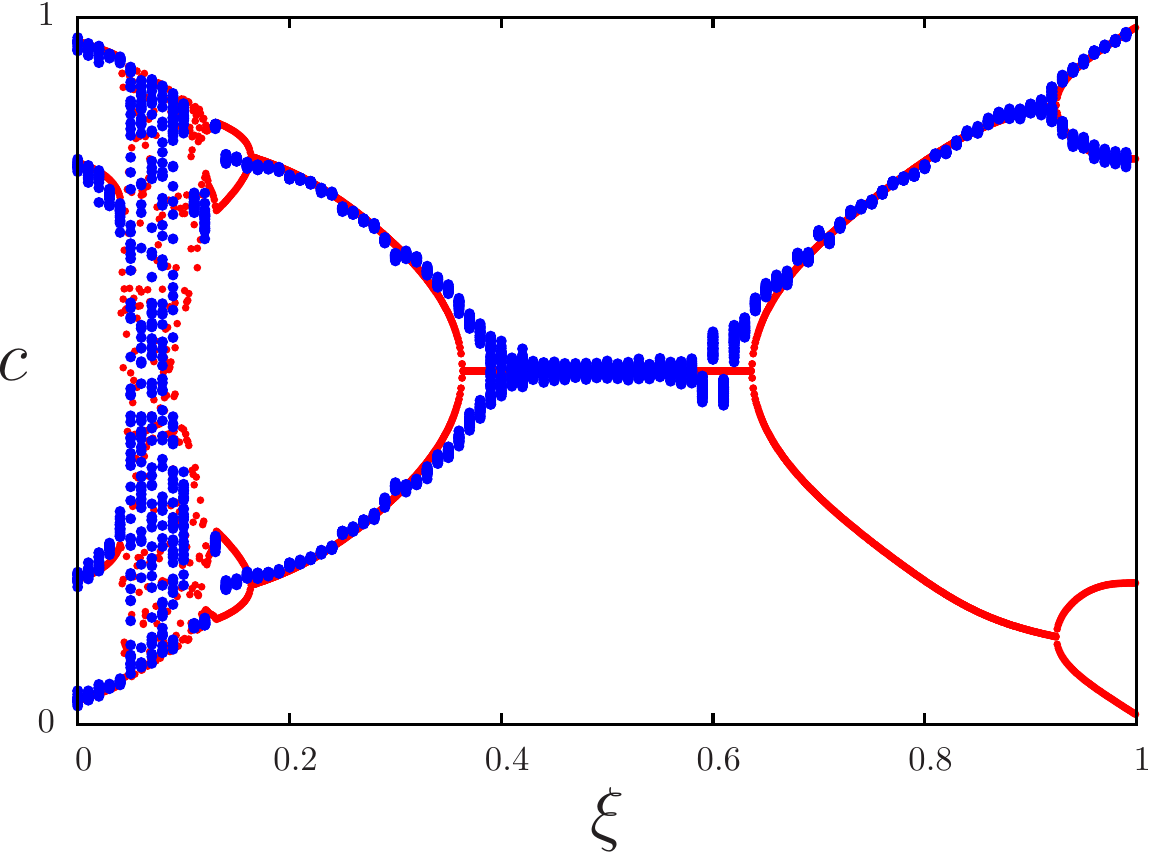} \\
  (c) & (d) \\
 \hskip -2mm
 \includegraphics[width=0.45\columnwidth]{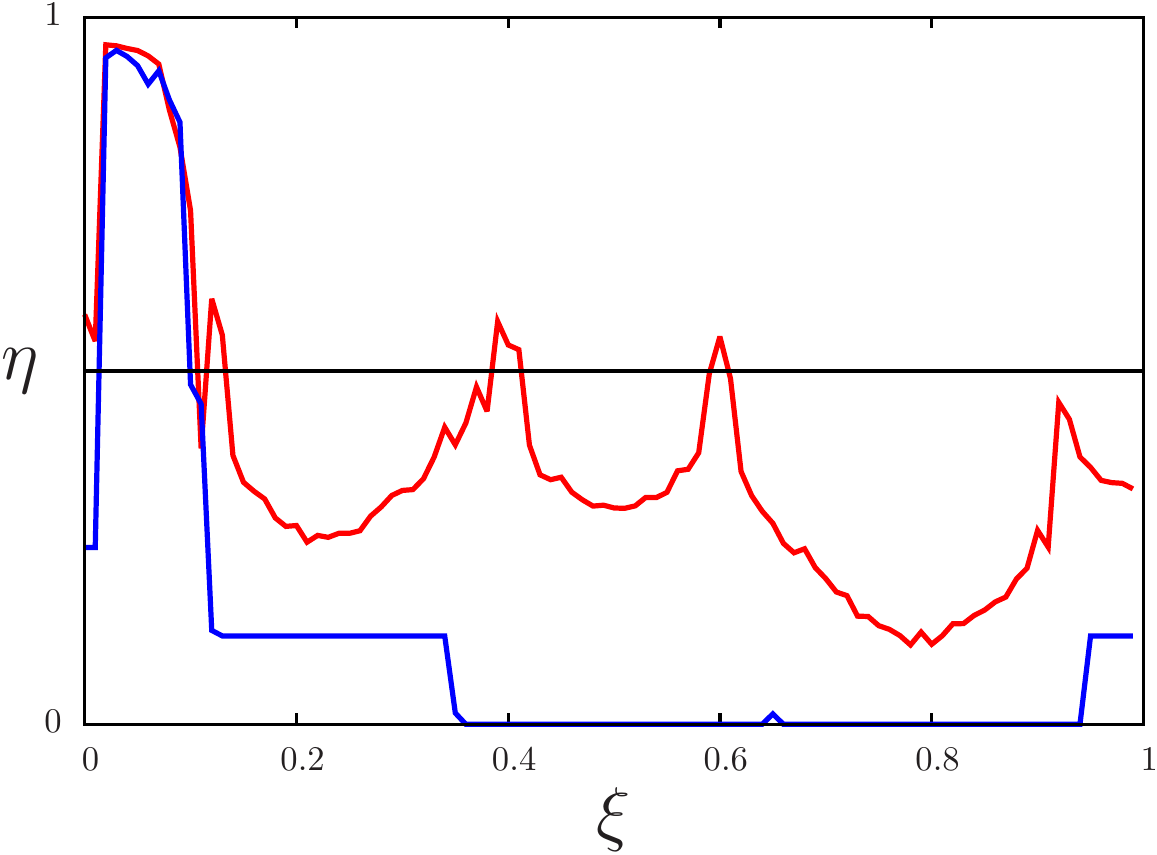} &
 \hskip 2mm
 \includegraphics[width=0.45\columnwidth]{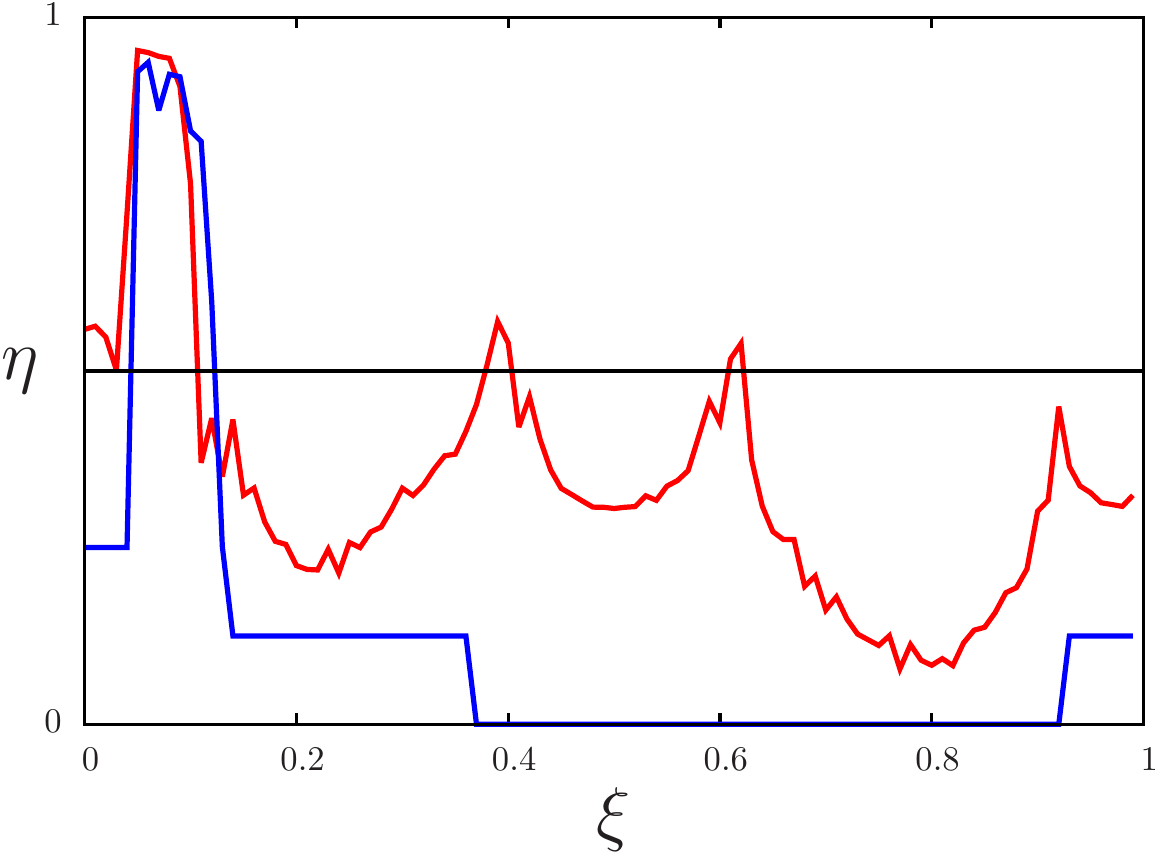} 
 \end{tabular}
 \end{center}
 \caption{\label{fig:sfn-mf} (a) and (b). Bifurcation diagrams of the
   average opinion $c$ as functions of the fraction of conformists
   $\xi$ on a scale-free network, large points (in blue) with $m=20$
   in (a) and $m=30$ in (b), and the mean field approximation,
   Eq.~(\ref{eq:mf}) with $k=1.7\times 20=34$ in (a) and $k=1.7\times
   30=51$ in (b), smaller points (in red). (c) and (d) Boltzmann's
   entropy $\eta$ of $c$ as functions of $\xi$ on a scale-free
   network, top curve for $\xi=1/2$ (in red), and of
   Eq.~(\ref{eq:mf}), bottom curve for $\xi=1/2$ (in blue).  (c)
   $m=20$, $k=34$. (d) $m=30$, $k=51$. In (a) and (b), for $c$ on a
   scale-free network and for each value of $\xi$, at $t=0$ the
   opinion of each agent is chosen at random in such a way that the
   average opinion is $c_0=0.9$ and for the mean field bifurcation
   diagram $c_0=9$ and $c_0=0.1$. After a transient of 300 time steps,
   the next 32 values of $c$ are plotted. For the  entropy $\eta$
   and each value of $\xi$, the unit interval is divided in $2^8=256$
   equal size subintervals and the frequency with which each
   subinterval is visited is found during $100\times 2^8=25,600$ time
   steps after a 300 time steps transient.}
\end{figure}

In Ref.~\cite{bagnoli2013} we showed that the dynamics of a model of a
society whose agents are all reasonable contrarians on a scale-free
network with $m$ initially connected agents is comparable to the
mean field approximation of Sec.~\ref{sec:meanfield} with connectivity
$k$ provided that
\begin{equation}
 \label{eq:alpha}
 k=\alpha m
\end{equation}
with $\alpha\sim 1.7$ for scale-free networks with $p(k)\propto
k^{-3}$. In Fig.~\ref{fig:sfn-mf} we show that this result also holds
for the model of societies studied here. In this Figure we compare the
bifurcation diagrams, as $\xi$ changes, of the dynamics on a
scale-free network with the corresponding mean field one,
Figs.~\ref{fig:sfn-mf} (a) and (b), and the scale-free entropies with
those of the mean field map, Figs.~\ref{fig:sfn-mf} (c) and (d).
There is a reasonable agreement in both bifurcation diagrams. Both
entropies show a good agreement where there is disorder, that is
$\eta>1/2$ but not when $\eta<1/2$. This can be understood from the
bifurcation diagrams. While the scale-free network dynamics is
stochastic and therefore the orbit visits many subintervals, the mean
field one visits a smaller number. For example, for $\xi=0.2$, the
mean field dynamics has period two so $\eta=2/8$ both in
Figs.~\ref{fig:sfn-mf} (c) and (d).


\section{Conclusions}\label{sec:conclusions}

The dynamics of the mean field approximation of the average opinion
$c$, Eq,~(\ref{eq:mf}) as a function of $\xi$ is chaotic with periodic
windows when $0<\xi<\xi_c$ and oscillates periodically between two
symmetric values when $\xi_c<\xi<\xi_a$.  For $\xi_a<\xi<\xi_b$,
$c=1/2$, and for $\xi_b<\xi$, $c>1/2$ ($c<1/2$) if $c_0=0.9$
($c_0=0.1$).  Also $\xi_a\sim 1-\xi_b$ and $\xi_a$ approaches a limit
value value as $k$ grows.  Eq,~(\ref{eq:mf}) depends on the
connectivity $k$ and on the transition probability $\tau$ that in turn
depends on $k$, $|J|$, $\eps$, and $q$. As far as we have explored the
parameter space, the above description is generic with the exception
that for large values of $k$ the chaotic phase is bounded below by
$\xi_d>0$. For $0<\xi<\xi_d$ there are two period two
orbits that depend on $c_0$ and for $\xi_d<\xi<\xi_c$ the orbits are chaotic
with periodic windows.

In small-world networks with small values of $p$,
Fig.~\ref{fig:sw-c-xi} (b), the coherent oscillations appear first for
a population with a small fraction of conformist rather than for a
pure-contrarian one. This is probably due to the fact that coherent
oscillations are a signal of synchronization, and the presence of
conformists increases the synchronization. On the other hand, if the
synchronized dynamics leads to a stable fixed point, the degree of
synchronization is not manifest. As a result, the first synchronized
zone is near the bifurcation point $\xi\simeq 0.$.

Dynamics on small-world networks approaches
that of the mean field approximation as the long range connection
probability $p$ grows. Dynamics of $c$ on scale-free networks is
similar to that of the mean fileld approximation provided $k=\alpha m$
with $\alpha\sim 1.7$. 


\section*{Acknowledgements} 
Interesting discussions with H\'ector D. Cort\'es Gonz\'alez and
Maximiliano Valdez Gonz\'alez are acknowledged.  This work was
partially supported by project PAPIIT-DGAPA-UNAM IN109213. F.B. acknowledges partial financial support from European Commission (ICT-20111.1.6) Proposal No. 288021 EINS - Network of Excellence in Internet Science and European Commission (FP7-ICT-2013-10) Proposal No. 611299 SciCafe 2.0.\\

We thank S. Galam and K.D. Harris for having pointed out their interesting contributions about the effects induced by the presence of contrarians and of chaotic bifurcations upon rewiring. 

\section*{Appendix} 

The bifurcation points $\xi_a$ and $\xi_b$ can be found from the mean
field evolution, Eq.~(\ref{eq:mf}), since at both values, the absolute
value of the derivative of this expression must be 1.  For large $k$ the
mean field approximation for the average opinion, can be approximated
by~\cite{bagnoli2013}
\begin{align}
 \label{eq:gauss}
c'=& \bigintsss dx \sqrt{\frac{k}{2\pi c(1-c)}}%
       \exp\left(-\frac{k(x-c)^2}{2c(1-c)}\right)\cdot\nonumber\\
   &\qquad\left[ \xi \tau(x;J) + (1-\xi)\tau(x;-J)   \right].
\end{align}

Expanding the right-hand side term of this expression around
$c=1/2$ up to first order
\begin{align}
 \label{eq:c'}
c'=& \sqrt{\frac{2k}{\pi}}\int dx\exp\left( -2k(x-c)^2\right)\cdot\nonumber\\
    &\qquad \left[ \xi \tau(x;J) + (1-\xi)\tau(x; -J)   \right]\nonumber\\
 =&\sqrt{\frac{2k}{\pi}} \int  dy \exp\left( -2ky^2\right)\cdot\nonumber\\
  &\qquad \left[ \xi \tau(y+c;J) + (1-\xi)\tau(y+c; -J)   \right].
\end{align}
We denote Eq.~(\ref{eq:c'}) by $g(\xi;J)$.

To proceed we need the derivative of $\tau$ given by Eq.~(\ref{eq:tau})
near $c=1/2$ and small $y$. From Eq.~(\ref{eq:tau})
\[
f(y; J)=\left.\frac{\partial \tau}{\partial c}\right|_{x=1/2+y}  =%
 \frac{4J \exp(-4Jy)}{(1+\exp(-4Jy))^2}.
\]
We note that $f(y;-J) = - f(y; J)$. Then
\[
g(\xi; J) = \bigintsss dy \sqrt{\frac{2k}{\pi}}%
   \exp\left( -2ky^2\right) (2\xi-1) f(y;J).
\]
It is now straightforward to check that 
\[
 g(1-\xi; J) = g(\xi; -J) = -g(\xi;J).
\]
If $g(\xi_a;J)=-1$ then $g(\xi_b)=1$ with $\xi_b=1-\xi_a$
and the the two bifurcation points are symmetric
with respect to $\xi=1/2$.

By approximating
\[
f(y;J) \simeq J(1-4 J^2 y^2) \simeq J \exp(-4J^2y^2)
\]
we get for $\xi_a$
\[
(2\xi_a-1)J\sqrt{\dfrac{k}{2J^2+k}} = -1
\]
i.e.,
\[
\xi_a(J, k) = \frac{1}{2}\left(1-\frac{1}{J}\sqrt{1+\frac{2J^2}{k}}\right).
\]
The limit of this last expression when $k\to\infty$ is $\xi_a(J,\infty))=(1/2)(1-1/J)$.


\begin{thebibliography}{99}



\bibitem{bagnoli2013} F. Bagnoli, R. Rechtman, Phys. Rev. E, {\bf 88}
 062914 (2013).

\bibitem{minority} W. B. Arthur, 
The American Economic Review
\textbf{84}, 406--411 (1994); D. Challet, Y.-C. Zhang, Physica A \textbf{246},  407--418 (1997)
D. Challet,  M. Marsili,  Y.-C. Zhang, Minority Games (Oxford University Press 2005);
A. C. C. Coolen, The Mathematical Theory of Minority Games
(Oxford University Press, 2005).


\bibitem{Ash} S.E. Asch,  \textit{Effects of group pressure on the modification and distortion of judgments}, in \textit{Groups, leadership and men}, H. Guetzkow editor  (Carnegie Press, Pittsburgh PA,  1951) pp. 177–190;
S.E. Asch,  \textit{Social psychology},  (Prentice Hall, Englewood Cliffs NJ, 1952); 
S.E. Asch, \textit{Studies of independence and conformity. A minority of one against a unanimous majority}. Psychological Monographs, \textbf{70}d, 1–70  (1956).

\bibitem{Galam-Contrarians} S. Galam, \emph{Contrarian deterministic effects on opinion dynamics: ``the hung elections scenario''}, Physica A \textbf{333}, 453 (2004).

\bibitem{Galam-Gemrev} S. Galam, \emph{Modeling the Forming of Public Opinion: An approach from Sociophysics},
Global Economics and Management Review \textbf{18} 11-20 (2013).

\bibitem{Galam-chaotic} C. Borghesi, S. Galam,  \emph{Chaotic, staggered, and polarized dynamics in opinion forming: The contrarian effect},  Phys. Rev. E \textbf{73}, 066118 (2006).

\bibitem{Dodds}
 P.S. Dodds, K.D. Harris, C.M. Danforth,
\emph{Limited Imitation Contagion on Random Networks: Chaos,
Universality, and Unpredictability},  Phys. Rev. Let. \textbf{110}, 158701 (2013).

\bibitem{Harris}  K.D. Harris, P.S. Dodds, C.M. Danforth,
\emph{Dynamical in uence processes on networks: General theory and
applications to social contagion}, Phys. Rev. E \textbf{88}, 022816 (2013).

\bibitem{Boltzmann} L. Boltzmann, {\em Vorlesungen \"uber Gastheorie}, Leipzig,
 J. A. Barth,; Partt I, 1896, Part II, 1898. English translation by S. G.
 Brush, {\em Lectures on Gas Theory}, University of California Press, 1964,
 Chapter I, Sec. 6.

\bibitem{Ott} E. Ott, {\em Chaos in Dynamical Systems}, Cambridge University Press (1993).

\bibitem{WattsStrogatz} D.J. Watts and S.H. Strogatz, Nature 
 \textbf{393}, 409 (1998).

\bibitem{Barabasi} L. Barab\'asi, R. Albert, Science {\bf 286} 509 (1999).

\bibitem{newman-PRE-64-026118-2001} M. E. J. Newman, S. H. Strogatz, D. J. Watts, Phys. Rev. E {\bf 64}, 026118 (2001).




\end{thebibliography}
\end{document}